\documentclass[twocolumn]{aa}

\usepackage[flushleft]{threeparttable}
\usepackage{graphicx}
\usepackage{amsmath}
\usepackage{subfigure}
\usepackage{amssymb}
\usepackage{tabularx}
\usepackage{natbib}
\usepackage{float} 
\usepackage{etoolbox}
\usepackage{amsmath}
\usepackage{color}
\usepackage{appendix}
\usepackage[usenames,dvipsnames]{xcolor}
\usepackage[normalem]{ulem}

\usepackage{multirow}

\begin{document}

\title{Addressing via N-body simulations the distribution \\ 
  of the satellite tidal debris in the Milky Way environment}
  
\author{Matteo Mazzarini\inst{1} \thanks{Matteo Mazzarini is a Fellow of the IMPRS-HD for Astronomy and Cosmic Physics.}
  \and Andreas Just\inst{1}
 \and Andrea V. Macci\`o\inst{2,3}
 \and Reza Moetazedian\inst{1}
 }

\offprints{M. Mazzarini, \email{mazzarini@uni-heidelberg.de}}

\institute{Zentrum f{\"u}r Astronomie der Universit\"at Heidelberg, Astronomisches Rechen-Institut, M{\"o}nchhofstr. 12-14, D-69120 Heidelberg, Germany
\and 
New York University Abu Dhabi, PO Box 129188, Abu
Dhabi, United Arab Emirates
\and
Max-Planck-Institut f{\"u}r Astronomie, K{\"o}nigstuhl 17,
D-69117 Heidelberg, Germany
}

\date{Received 23 January 2020 / Accepted 12 March 2020}

  
  \titlerunning{Distribution of the MW satellite debris}
  \authorrunning{Mazzarini, M., Just, M., Macci\`o, A. V., Moetazedian, R.}
  \abstract
   {} 
   {We study the distribution of the Milky Way satellites stellar and dark matter debris.}
   {For the first time we address the question of the tidal disruption of satellites in simulations by utilising simultaneously a) a realistic set of orbits extracted from cosmological simulations, b) a three component host galaxy with live halo, disc and bulge components, and c) satellites from hydrodynamical simulations. 
We analyse the statistical properties of the satellite debris of all massive galaxies 
   reaching the inner Milky Way on a timescale of 2\,Gyr.
   }
   {Up to 80\% of the dark matter is stripped from the satellites, while this happens for up to 30\% of their stars. The stellar debris ends mostly in the inner Milky Way halo, whereas the dark matter debris shows a flat mass distribution over the full main halo. The dark matter debris follows a density profile with inner power law index $\alpha_{\rm DM}=-0.66$ and outer index $\beta_{\rm DM}=2.94$, while for stars $\alpha_{*}=-0.44$ and $\beta_{*}=6.17$. In the inner 25\,kpc, the distribution of the stellar debris is flatter than that of the dark matter debris and the orientations of their short axes differ significantly.
   Changing the orientation of the stellar disc by 90$^{\rm{o}}$ has only a minor impact on the distribution of the satellite debris.
   }
   {Our results indicate that the dark matter is more easily stripped than stars from the Milky Way satellites. The structure of the debris is dominated by the satellite orbital properties.
   The radial profiles, the flattening and the orientation of the stellar and dark matter debris are significantly different, which prevents the prediction of the dark matter distribution from the observed stellar component.}

\keywords{methods: numerical --  Galaxy: kinematics and dynamics -- (galaxies:) Local Group} 
\maketitle 

\section{Introduction}

Within the $\Lambda$-Cold Dark Matter ($\Lambda$CDM) theoretical framework \citep{WhiteRees78} it is expected that the Milky Way (MW) Galaxy is surrounded by several satellite galaxies. The $\Lambda$CDM framework supports the hierarchical scenario of structure formation, according to which smaller dark matter (DM) haloes form in the earlier stages of the Universe, and later merge to form higher mass structures, with the central, massive host galaxies being the end product of this process \citep{Blumenthal84}. The satellite galaxies around the MW are the remnants of its past accretors.

A stream of gas, the so-called Magellanic Stream, is observed in the MW environment \citep[]{WannierWrixon72,Mathewson74,Nidever10}. 
This stream is attributed to the tidal interaction between the Large and Small Magellanic Clouds (LMC and SMC), with a possible contribution of the MW environment to this process \citep{DiazBekki11,Besla16,DonghiaFox16, Wang19}.   
Observations of the northern Galactic hemisphere have evidenced the presence of stellar overdensities in the shape of two elongated streams \citep{Ibata94,Newberg07}. It is believed that these streams are the debris of a progenitor (identified with the Sagittarius dwarf galaxy) that has been interacting with the MW in the last Gyrs and thereby has dissolved in the environment, losing matter \citep{Martinez-Delgado04}.
Other observations include the discovery of 
the Monoceros-Canis Major streams, with a progenitor of estimated mass equal to the one of the Sagittarius dwarf, i.e. around $10^9\,$M$_{\odot}$ \citep{Newberg02}.

Not only streams from dwarf galaxies currently being destroyed are observed in the MW environment, but signs of the past accretion of satellites can also be detected. 
From the abundance analysis of Gaia Data Release 2 \citep[DR2,][]{Brown18}, recent work from \cite{Helmi18} suggests that most of the inner halo of our Galaxy originated from a past, major impact with the so-called Gaia Enceladus progenitor around 10 Gyr ago.  
Using DR2 photometry data combined with spectroscopic information from other surveys \citep{Wilson10,Cui12,Kunder17,Abolfathi18,Marrese19}, \cite{Koppelman19} identified new members of the Helmi stellar stream, originally identified in the solar neighborhood by \citep{Helmi99}. Combining photometry and metallicity measurements with N-body experiments, they found low metallicity populations and predicted that the Helmi stream originates from a dwarf galaxy of mass $M\sim10^8 $M$_{\odot}$, accreted onto the MW 5 Gyr ago.

Different observational and theoretical studies have focused on the spatial distribution of the MW satellites and of their debris, evidencing the presence of the so-called Vaste Plane of Satellites (VPOS) but also aligned streams of disrupted satellites \citep{Pawlowski12}. By means of numerical simulations, \cite{SantosSantos19} found that the inertia tensor related to the spatial distribution of the bulk of satellites around the MW has a flatness $c/a$ ranging from 0.1 for less than 10 satellites around the MW to 0.2 for more than 25 satellites around the MW. \cite{Buck16} showed however that even when forming these spatially coherent planes, satellites seem to be kinematically incoherent and these planar structures do not last a long time.   
\cite{Lisanti&Spergel12} focused on the DM debris in the Via Lactea II simulation and found evidence for spatial homogeneity of its distribution.
\textcolor{black}{These results seem contrasting, in the perspective of understanding the distribution of the MW satellites and of their stellar and DM debris.}

When considering the origin of the MW halo, the theoretical work of \citep{Pillepich15} predicted that $70 \%$ of MW stellar content is formed in-situ, that the majority of the ex-situ stars come from infalling satellites and characterise most of the stellar halo, and that the mass of these accreting satellites is relevant to the formation process of the MW halo. The additional study of \citep{Deason16} suggested that the main contribution to the MW halo build-up (around $10^9$ M$_{\odot}$) comes from larger satellites such as LMC and SMC, whereby the contribution from ultra-faint dwarf galaxies (UFDs, with stellar masses below $10^5$ M$_{\odot}$) is minimal. 

Large numerical investigations such as the Aquarius \citep[Aq, ][]{Springel08} and Via Lactea II \citep{Diemand08} cosmological simulations successfully described the clustering properties of Cold DM in large portions of the  Universe ($\sim 100$ Mpc) down to the scales of subhaloes and the main MW-halo core ($\sim 10-100$ pc). These projects made use of state-of-the-art gravitational solvers \citep[\textsc{Gadget 3} after \textsc{Gadget 2}, and \textsc{pkdgrav}; see][]{Springel05,Stadel01} but lacked prescriptions for baryonic physics, which affects the final properties of the galactic haloes \citep{PontzenGovernato12} and in the case of galaxies like the MW influences the process of matter stripping from their satellite galaxies \citep{Garrison-Kimmel17}. More recently, big simulation projects were dedicated to combine gravity and 

baryonic physics with refined numerical recipes \citep[see \textsc{Arepo},][]{Springel10} in order to obtain more accurate results, in better agreement with observations of galaxies in the Universe \citep{Vogelsberger14,Vogelsberger14b,Genel14,Sijacki14} or more specifically with the observational properties of dwarf galaxies \citep[Latte simulation]{Wetzel16}. 

The zoom-in simulations Auriga \citep{Grand17} and Eris \citep{Guedes11} focus on the detailed properties of MW-like objects with a higher resolution ($6 \times 10^{3}$ M$_{\odot}$ for gas particles and $4 \times 10^4$ M$_{\odot}$ for DM in Auriga, and $9.8 \times 10^4$  M$_{\odot}$, $2 \times 10^4$ M$_{\odot}$ and $6 \times 10^3$ M$_{\odot}$ for DM, gas and stars in Eris). For their study on the environmental effects on MW satellites, \cite{Buck19} adopted a similar mass resolution, with stellar particles reaching a resolution of $6.7 \times 10^3$ M$_{\odot}$ and DM particles having a resolution of up to $\sim 10^5$ M$_{\odot}$. For comparison, in the IllustrisTNG (The Next Generation) project, the typical resolution of particles in TNG50 \citep[its highest resolution simulation box,][]{Pillepich19,Nelson19} 
is $8.5 \times 10^{4}$ M$_{\odot}$ for stars and $4.5 \times 10^{5}$ M$_{\odot}$ for DM. In other projects like \textsc{Apostle} \citep[A Project Of Simulating The Local Environment,][]{Sawala16,Sawala17}, which focuses on reproducing the kinematics and dynamics of the Local Group, the best resolutions for stellar and DM particles are $10^4$ M$_{\odot}$ and $5 \times 10^4$ M$_{\odot}$, respectively.

While keeping in line with the best available resolution of DM particles ($M = 3.4 \times 10^3$ M$_{\odot}$ at best), \cite{Wang15} made an advancement in resolving baryons for their \textsc{NIHAO} (Numerical Investigation of a Hundred Astrophysical
Objects) cosmological simulations, with adopted masses as small as $M = 6.2 \times 10^2$ M$_{\odot}$ for gas particles. Their high-resolution simulations were used to study the evolution and properties of galaxies in a range of masses going from dwarf DM haloes of mass $M \sim 10^9$ M$_{\odot}$ to MW-like halos with masses $M \sim 10^{12}$ M$_{\odot}$.
Higher resolutions were also employed in cosmological simulations of isolated dwarf galaxies by \citet[][hereafter M17]{Maccio17}, using the N-body code \textsc{Gasoline} \citep{Wadsley04} with Smoothed particle hydrodynamics (SPH) prescriptions for gas dynamics \citep{Lucy77, GingoldMonaghan77}. For their dwarf galaxy models, M17 reached resolutions as high as $M = 1.2 \times 10^2$ M$_{\odot}$, $M = 4 \times 10$ M$_{\odot}$ and $M = 6 \times 10^2$ M$_{\odot}$ for gas, star and DM particles respectively. 
M17 evolved their dwarf galaxy models in isolation (i.e., without any host MW  exerting gravitational/hydrodynamical effects on them) in the redshift range [$100<z<1$]. The position-velocity (i.e., phase-space) distribution of stars and DM at $z=1$ in these objects reflects the additional hydrodynamics, gas cooling and stellar feedback recipes implemented in the M17 simulations. Having switched on Star Formation (SF) in their simulations, M17 obtained satellites consisting of a realistic combination of DM particles, gas particles and star particles. 

Given the above picture, in this work we present our study on the general properties of the MW satellites debris via numerical simulations. To do this, we adopted a hybrid approach, for which we combined high resolution MW models from previous literature and with parameters extracted from cosmological data, with M17 high resolution dwarf galaxy models (employing them as satellites of the MW). Furthermore, the initial distribution of the satellites around the MW comes from the results of the Aq cosmological simulations. 

Our approach is similar to what was employed in \citet[hereafter MJ16]{MoetazedianJust16}. Following the numerical prescriptions of \cite{YurinSpringel14}, MJ16 combined cosmological initial setup and high resolution numerical simulations. They modeled 6 MW numerical realisations, each consisting of live disc, bulge and halo, and each with halo parameters extracted from the Aq A2-to-F2 cosmological simulations at redshift $z=0$. They combined each MW model with a number of N-body satellites for which they extracted the parameters from the corresponding Aq snapshot.
The mass resolution of their discs is $3.4 \times 10^3$ M$_{\odot}$, higher than the stellar resolution in cosmological simulations. With this hybrid approach, they were able to study the effect of a cosmologically motivated set of satellites on a high-resolution MW disc.

Since we wanted to address the distribution of both the stellar and DM satellite debris in our study, in contrast to MJ16 we decided to employ a selection of satellites made of baryons and DM (hybrid satellites hereafter), that we extracted from the sample of dwarf galaxies described and studied in M17.

This paper is organised as follows: in Section 2 we show the selection of satellite models for our simulations and we discuss the properties of the satellites. At the end of Section 2 we show the numerical setup of our simulations. In Section 3 we show our results, addressing the stripping of satellite matter, the radial distribution of the debris and its shape. In Section 4 we additionally investigate the surviving fraction of satellites at the end of our simulations and the final DM/stellar mass ratio inside the surviving satellites. 
In Section 5 we conclude and we discuss our results. 

\section{Numerical simulations}

We ran a set of N-body simulations of MW-satellites interaction to address the properties of the stripped satellite debris.
For each simulation, we took the MW model used by MJ16 for the corresponding Aq setup. 
The initial MW data were originally extracted by MJ16 as Navarro-Frenk-White (NFW) haloes \citep{Navarro95b} and with a best-matching numerical profile, with total mass equal to the NFW virial mass $M_{200}$ \cite[i.e., the mass enclosed in the radius with average inside density equal to 200 times the average cosmic density,][]{Navarro95} of the original NFW model and with the same inner density profile, following the prescriptions of \cite{Springel05b}. 
The discs of these numerical models have an exponential profile with the distance of the Sun set to $R_{\rm{0}} = 8$ kpc from the Galactic centre, while the bulge has a Hernquist density profile \citep[see][]{Hernquist90}.     

For each MW halo selected in the corresponding Aq simulation, MJ16 employed a subhalo mass cut of 
$10^8$ M$_{\odot}$ and required the subhaloes to have had a pericentre passage within 25 kpc of the MW within 2\,Gyr. They resimulated these systems as higher-resolution DM-only N-body spheroids (50K particles each), for a total number of satellites per simulation ranging from a minimum of 12 (Aq-E2) to a maximum of 24 (Aq-F2). Their satellites have a range in mass that spans from $10^8 \rm{M}_{\odot}$ \footnote{The reason for this mass cut in MJ16 is because they addressed the impact of satellite galaxies on the MW disc kinematics and dynamics and in this case they show that the impact of satellites increases with the satellite mass as $\propto M^2$.} to $6 \times 10^{10} \rm{M}_{\odot}$. They modeled each satellite as an NFW profile, and each satellite has equal mass particles. They placed the satellites in the respective positions indicated from the corresponding Aq simulations. 
In Table \ref{tab:setup_MW} we show the main properties of the MW models.

\begin{table*}[ht!]
\caption{\footnotesize Simulation setup for the MW models employed in the MJ16 simulations and the corresponding number of satellites. Left to right columns: Aq run, number of MW disc particles, bulge particles and halo particles, number of satellites in the simulation, mass of disc, mass of bulge, virial mass of the halo, halo NFW scale radius.} 
\centering      
\begin{tabular}{c| c c c c c c c c}          
\hline\hline                        
Aq run & $N_{\rm{disc}}$ & $N_{\rm{bulge}}$ & $N_{\rm{halo}}$ & $N_{\rm{sat}}$ & $M_{\rm{disc}}$ & $M_{\rm{bulge}}$ & $M_{200}$ & $r_{\rm{scale,NFW}}$ \\
& & & & & (M$_{\odot}$) & (M$_{\odot}$) & (M$_{\odot}$) & (kpc)\\
\hline            

    A2  & \multirow{6}{*}{$1\times10^7$} & \multirow{6}{*}{$5\times10^5$} & \multirow{6}{*}{$4\times10^6$} & 20 & \multirow{6}{*}{$3.4\times10^{10}$} & \multirow{6}{*}{$0.9\times10^{10}$} & \multirow{6}{*}{$1.77\times10^{12}$} & 15.00\\ 
    B2  & & & & 17 & & & & 24.98\\      
    C2  & & & & 14 & & & & 15.96\\
    D2  & & & & 23 & & & & 25.91\\
    E2  & & & & 12 & & & & 29.39\\
    F2  & & & & 24 & & & & 24.80\\
\hline 
\end{tabular}
\label{tab:setup_MW}  
\end{table*}

\subsection{Selecting the satellite galaxies}
\label{ssec:select_sats}

For the selection of the satellite galaxies to be used in our simulations in place of the DM-only ones from MJ16, we extracted the best dwarf galaxies from the sample of M17. 
 
The two samples of satellites come from different simulations (DM-only versus full N-body-SPH) run up to different final redshifts ($z_{\rm{final}}$=0 versus $z_{\rm{final}}$=1). Therefore, we chose to match them by minimizing the distances of the two satellite samples in the  $\log(M_{\rm{200}})$-$\log\Big[(v_{\rm{max}}/r_{\rm{max}})^2\Big]$ space. Here, 
$v_{\rm{max}}/r_{\rm{max}}$ is the ratio between the maximal circular velocity of the satellite and the radius of maximal circular velocity. This last quantity is an indicator of the inner density (and therefore of the depth of the potential well) of each satellite. In fact,

\begin{equation}
\begin{split}
    \Bigg(\frac{v_{\rm{max}}}{r_{\rm{max}}}\Bigg)^2 = \frac{G \times M(<r_{\rm{max}})}{r_{\rm{max}}} \times\frac{1}{r_{\rm{max}}^2} =  \frac{4 \pi}{3} \bar{\rho}(<r_{\rm{max}}) \, ,
\end{split}
\end{equation}
where $\bar{\rho}(<r_{\rm{max}})$ is the average density within the radius of maximal circular velocity.

\begin{figure}[thb]
  \centering
  \includegraphics[scale=0.46]{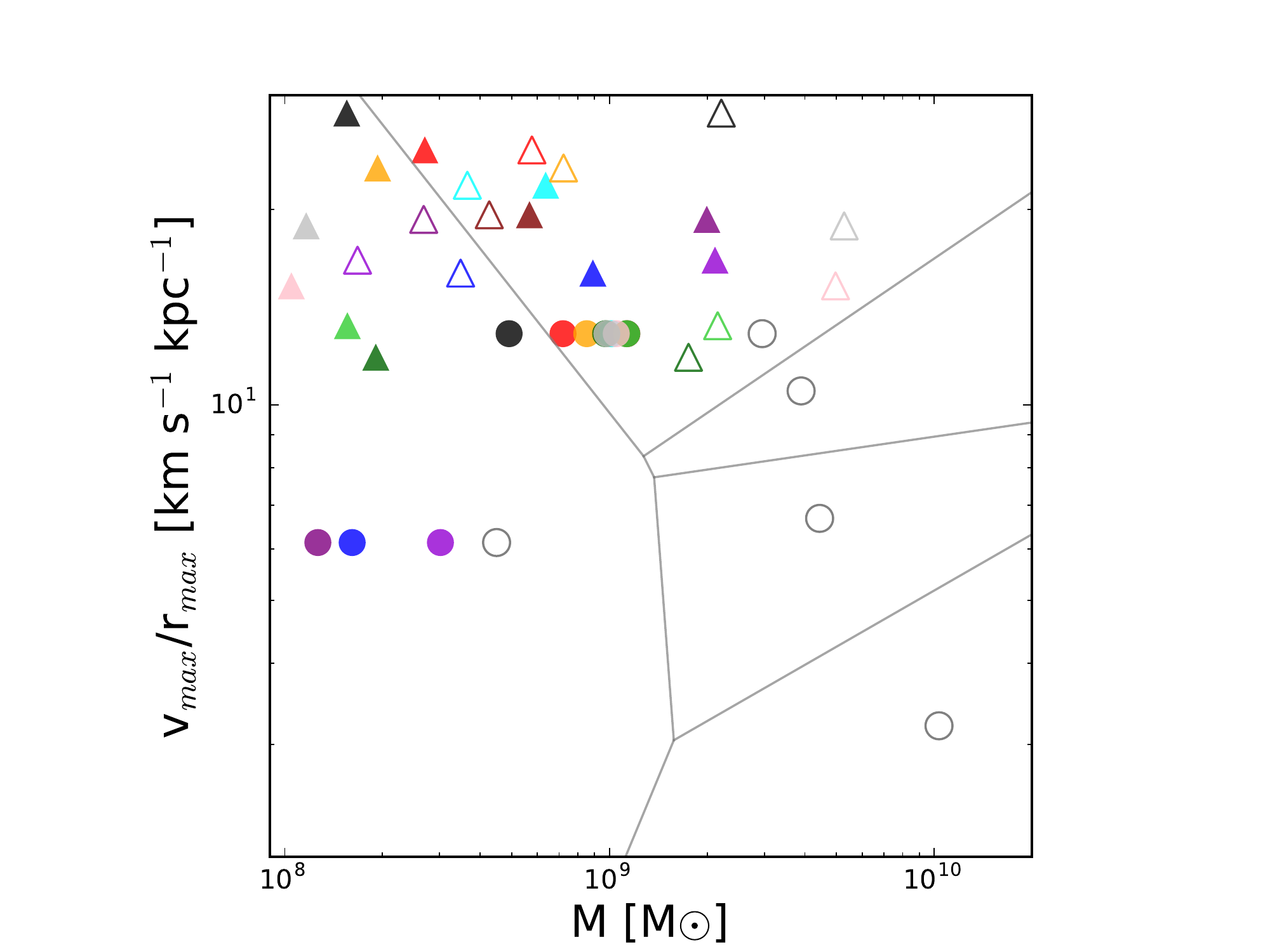}
  \caption{Distribution of the candidate best-matching dwarf galaxies in the $\log(M_{\rm{200}})-\log\left(v_{\rm{max}}/r_{\rm{max}}\right)$ plane. Empty grey circles represent all the 5 best-matching dwarf galaxies of the M17 sample, which are used for the six Aq-simulations (for their properties see Table \ref{tab:setup_5_sats}). The tessellation with grey segments represents the division in regions where the satellite galaxies of MJ16 are closest to any of the 5 best matching dwarf galaxies of M17. For the case of Aq-E2 the satellites of MJ16 are shown as empty triangles according to their $M_{\rm{200}}$ and $v_{\rm{max}}/r_{\rm{max}}$ values. Filled triangles are the same satellites after tidal cutting, i.e. with their $M_{\rm {tid}}$ values. The filled circles show the corresponding M17 satellites after tidal cutting. The colours of the filled symbols represent the matched pairs. For Aq-E2 only sat1 and sat4 have triangles falling in their regions. }
\label{fig:1}
\end{figure}
Due to the intrinsic differences in properties of the two samples of objects, we do not expect the matching sample to be as homogeneously distributed in the log-log space as the MJ16 satellites. Consequently, 5 of the M17 satellites fall in the relevant parameter range and so the candidate dwarfs that match our constraints are repeated for most times. We also note that the selected objects tend to be lower in central density than the ones from MJ16. In Table \ref{tab:setup_5_sats} we show the number of particles and the masses of the gas, DM and stellar component for each of the 5 selected dwarfs. 

In the next step we cut the total masses of the selected dwarfs to their initial tidal radii.
In order to do this mass cut, we first calculated the tidal radius of each satellite of MJ16, as in \cite{ErnstJust13}, 
\begin{equation}
\label{eq:tidal_rad_definition}
r_{\rm{tid}}(r) = \left(\frac{M_{\rm tid}}{{\omega}^2 -\frac{d^2\Phi(r)}{dr^2}}\right)^{\frac{1}{3}} \, ,
\end{equation} 
where $M_{\rm{tid}}$ is the tidal mass of the satellite, $r$ is its Galactocentric distance (GCd), 
$\omega$ is its orbital angular speed and $\Phi$ is the gravitational potential of the MW.
In order to obtain the tidal radius for each M17 satellite, we used the approximation
\begin{equation}
    r_{\rm{tid,M17}} = r_{\rm{tid,MJ16}} \Bigg(\frac{M_{\rm{200,M17}}}{M_{\rm{200,MJ16}}} \Bigg)^{\frac{1}{3}} \, . 
\label{eq:tidal_rescaling}    
\end{equation}
 The resulting distribution of satellites masses are shown as full coloured symbols for the case of Aq-E2 in Figure \ref{fig:1}. Since $v_{\rm{max}}$ and $r_{\rm{max}}$ are not altered by the tidal cutting, the satellites are shifted horizontally in the figure. For Aq-E2 only the two satellites sat1 and sat4 of M17 were relevant. 
\begin{table*}[t!]
\caption{\footnotesize Properties of the 5 selected satellites, before they are cut in tidal radii. The satellites are ordered from sat1 to sat5 according to decreasing total DM mass. Left to right columns: satellite name, number of gas particles, number of DM particles, number of star particles, total gas mass, total DM mass, total stellar mass, $v_{\rm{max}}/r_{\rm{max}}$ and $r_{\rm{max}}$.}

\centering      
\begin{tabular}{c| c c c c c c c c}          
\hline\hline                        
 \\
Sat & $N_{\rm{gas}}$ & $N_{\rm{DM}}$ & $N_{*}$ & $M_{\rm{gas}}$ & $M_{\rm{DM}}$ & $M_{*}$ & $v_{\rm{max}}/r_{\rm{max}}$ & $r_{\rm{max}}$\\
& & & & (M$_{\odot}$) & (M$_{\odot}$) & (M$_{\odot}$) & (km s$^{-1}$ kpc$^{-1}$) & (kpc)\\
\hline                          
    sat1  & $3.5\times10^5$ & $3.7\times10^6$ & $8.2\times10^4$ & $1.92\times10^8$ & $1.02\times10^{10}$ & $8.97\times10^6$ & 3.34 & 12.35 \\ 
    sat2  & $1.20\times10^5$ & $1.05\times10^6$ & 7880 & $9.80\times10^7$ & $4.34\times10^9$ & $1.27\times10^6$ & 6.37 & 4.65\\ 
    sat3  & $5.99\times10^4$ & $9.32\times10^5$ & 8116 & $4.92\times10^7$ & $3.85\times10^9$ & $1.30\times10^6$ & 10.37 & 3.11\\ 
    sat4  & $6.54\times10^4$ & $1.06\times10^6$ & 5194 & $3.61\times10^7$ & $2.92\times10^9$ & $5.46\times10^5$ & 12.27 & 2.53 \\ 
    sat5  & 1764 & $1.62\times10^5$ & 406 & $9.72\times10^5$ & $4.49\times10^8$ & $4.25\times10^4$ & 6.26 & 2.40 \\ 
    
\hline 
\end{tabular}
\label{tab:setup_5_sats}  
\end{table*}

\subsection{Numerical dwarf galaxies as candidate satellites: properties}

We want to show in this section that, thanks to our hybrid approach, we achieve a mass resolution which is an order of magnitude better than what is possible in current self-consistent cosmological simulations; that the density profiles of their stellar and DM components have different slopes, thereby allowing us to treat them as two distinct populations; and that the mass distribution of DM and stars as a function of specific energy returns two distinct populations that one cannot naturally derive from the DM-only satellites.

To do this, we show the particle mass distribution, the density profile and the mass distribution as a function of specific energy of satellite 3 as representative for the 5 satellites, since they all have similarities in these aspects.

\subsubsection{Mass resolution}

\begin{figure}
  \centering 
  \includegraphics[scale=0.39]{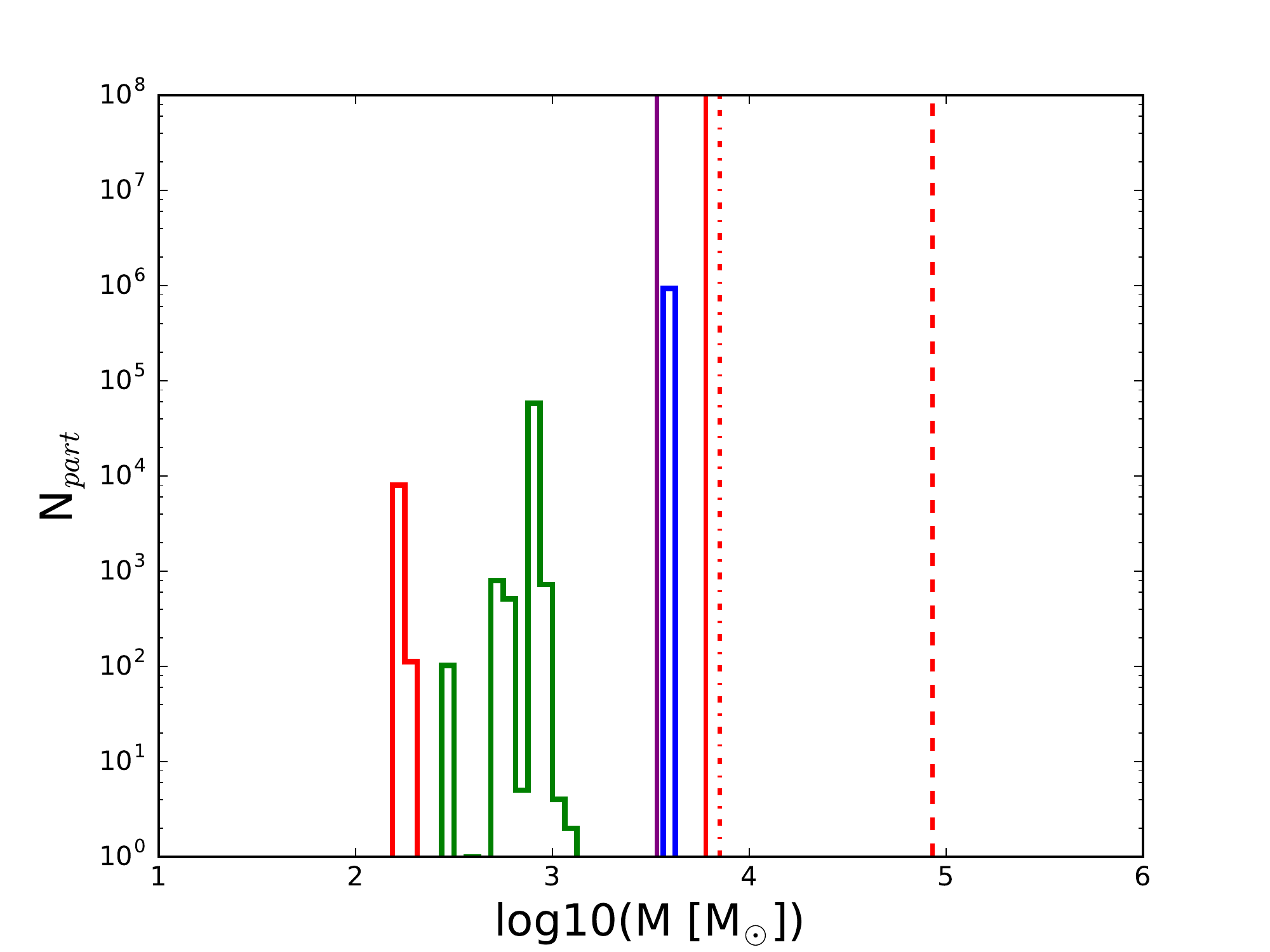}\\
  \includegraphics[scale=0.39]{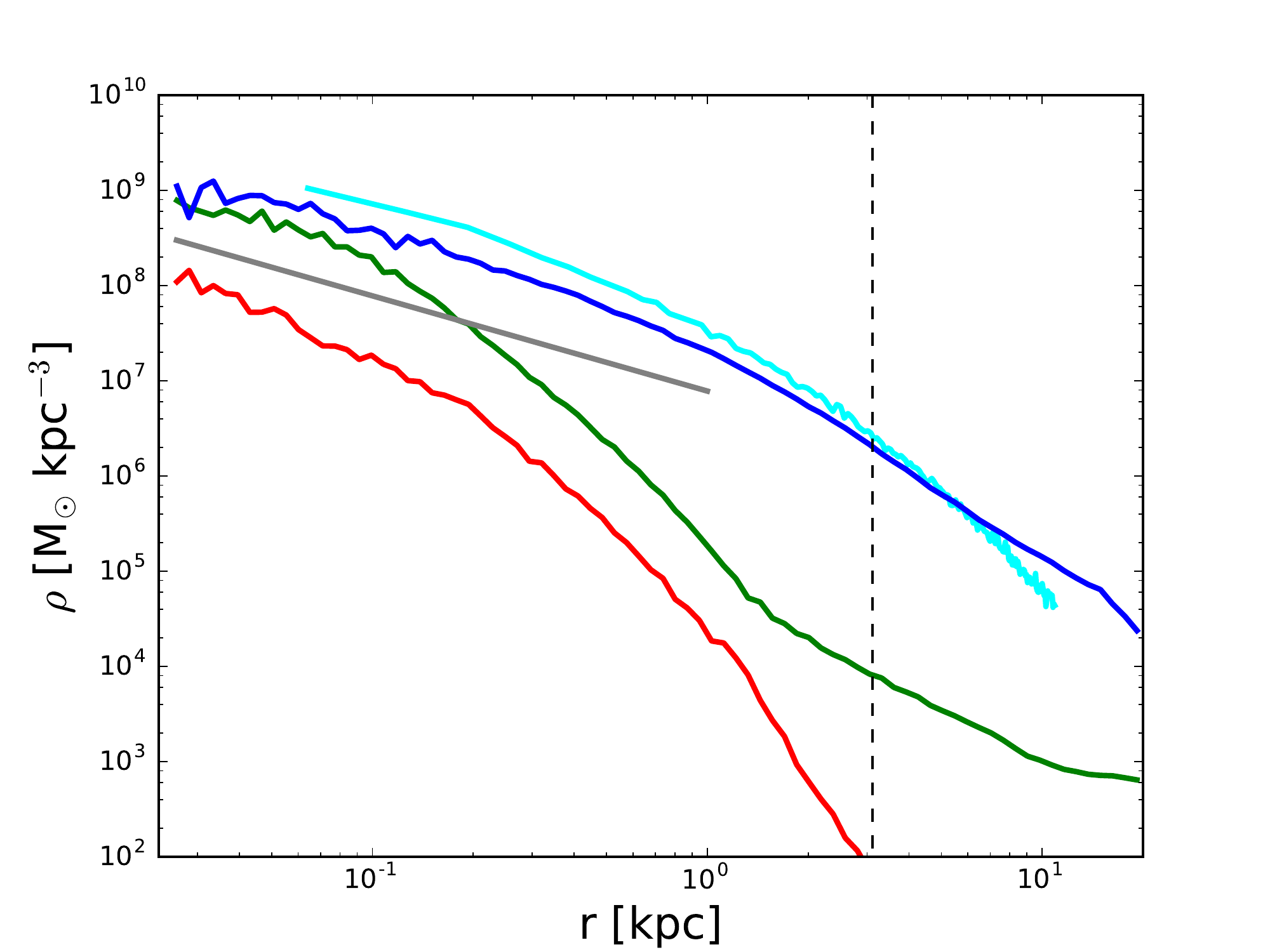}\\
  \includegraphics[scale=0.39]{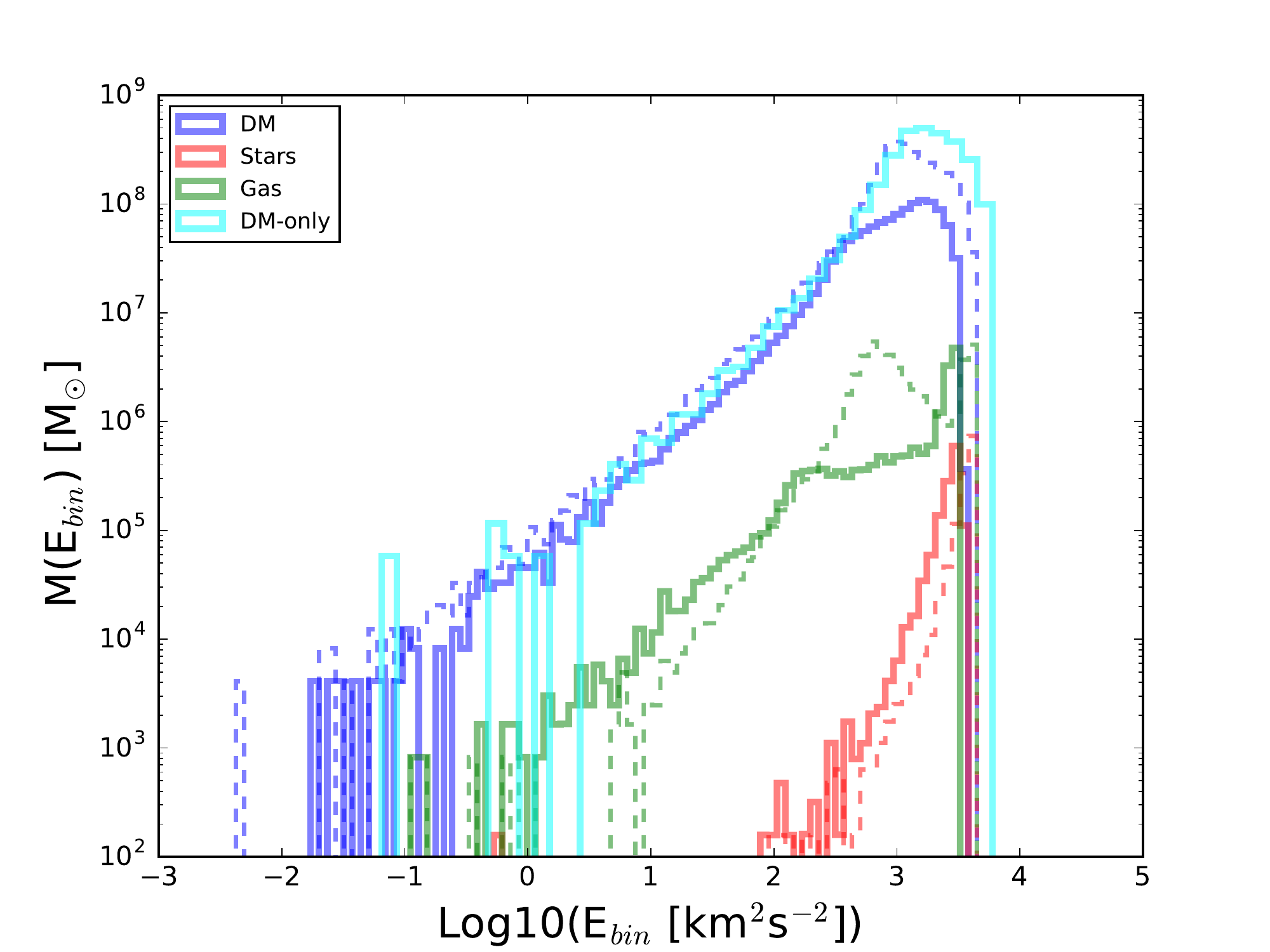}
  \caption{\footnotesize From top to bottom, the particle mass distribution, the radial density profiles and the specific energy distributions for the 3 components (stars, gas and DM) of satellite 3. In all figures, red is for stars, green for gas and blue for DM. Cyan is for one
  corresponding DM-only satellite from MJ16. \textit{Top panel}: the additional vertical dashed red line is the stellar mass resolution for the TNG50 simulation, the red dot-dashed vertical line is for Latte and the red thick line is for Eris. The purple thick line is for the disc resolution in the MW model of MJ16. \textit{Central panel}: the density profiles are given as a function of the distance $r$ from the density centre of the satellite. For comparison, the thick grey line represents an NFW inner profile (with radial dependence $\propto r^{-1}$). The black dashed vertical line marks the tidal radius position of satellite 3 after tidal radius rescaling from the corresponding DM-only satellite. \textit{Bottom panel}: the dashed and thick histograms represent the specific energy distribution in terms of mass per energy bin for each component of satellite 3 before and after applying the tidal cut to the satellite, respectively. The cyan line shows the same for the DM only satellite.}
  \label{fig:sat3MassesDensprofEnergy}
\end{figure}

In the top panel of Figure \ref{fig:sat3MassesDensprofEnergy} we can see that DM and stellar particle masses are orders of magnitude below the masses of the corresponding particles in the cosmological simulations TNG50, Eris and Latte. The disc of our MW models, which is made of collisionless stellar particles, has a better resolution than the stars in the Eris simulation. 
This puts our satellite models as well as the disc models in a better position in terms of resolution. The fact that the stellar and DM debris particle masses are at least 2 orders of magnitude below the corresponding, best cosmological simulation resolutions allows for a more accurate statistical investigation of the stellar and DM debris in our simulations, avoiding low-number noise in the calculations.  

\subsubsection{Density profiles} 

The central panel of Figure \ref{fig:sat3MassesDensprofEnergy} shows the density profiles of gas, DM and stars in satellite 3, as well as the density profile of a DM-only satellite that is substituted by satellite 3 in Aq-B2. 

The grey straight line shows the inner slope  $\propto r^{-1}$ of a NFW profile for comparison. We can see that while DM and gas have a slope of their inner density profiles close to the NFW case,
the stellar component has a significantly steeper inner density profile. This is already a hint that there is no simple recipe to select a realistic stellar component based on a DM only simulation.

\subsubsection{Specific energy distribution}

For each satellite particle, we calculated its total specific binding energy (i.e., binding energy per unit mass) as

\begin{equation}
    E_{\rm{bin}} = - \epsilon_{\rm{tot}} \quad ,
\end{equation}
with

\begin{equation}
    \epsilon_{\rm{tot}} = \epsilon_{\rm{pot}} + \epsilon_{\rm{kin}} \quad ,
\end{equation}
where $\epsilon_{\rm{tot}}$, $\epsilon_{\rm{pot}}$ and $\epsilon_{\rm{kin}}$ are the specific total, potential and kinetic energy of the particles.
For the gas particles, an additional term was counted in the sum, $\epsilon_{\rm{therm}}$, the specific thermal energy, calculated as
\begin{equation}
    \epsilon_{\rm{therm}} = \frac{3}{2} \frac{n}{\rho} k_{\rm{B}} T \quad ,
\end{equation}
where $k_{\rm{B}}$ is the Boltzmann constant and 

\begin{equation}
    \frac{n}{\rho} = \frac{1}{0.6 m_{\rm{H}}}
\end{equation}
is the number of particles per unit mass for fully ionised gas, with the term $m_{\rm{H}}$ at the denominator being the hydrogen mass, $m_{\rm{H}} = 1.67 \times 10^{-27}$ kg. 

Since the particle masses of the M17 satellites vary over a large range, we weighted the particle distributions in specific energy by their masses. 
In the bottom panel of Figure \ref{fig:sat3MassesDensprofEnergy} we plot the specific energy distributions of each component for the satellite 3 at $z = 1$ before and after tidal cutting. 
For comparison the specific energy distribution function (also weighted by particle mass) of a corresponding DM-only satellite of MJ16 is shown in cyan.
Even after tidal cutting, the gas and DM particles of M17 dwarfs and the DM particles of MJ16 satellites have specific energy distribution functions with a similar slope. In contrast, the shape of the distribution function of the stellar component is very different to the distribution of the DM component of the M17 and the MJ16 satellites. This confirms the need for baryonic physics in the satellite models in order to obtain a realistic  
stellar component.
 
\subsection{Numerical setup and simulation properties}

We run a total of 12 simulations, two for each corresponding Aq setup from MJ16. 
For the first set of 6 simulations, one for each Aq setup from MJ16, we put the 5 best-matching satellites in place of the corresponding MJ16 satellites at the same initial positions and velocities. 
For the second set of 6 simulations, again one for each Aq setup from MJ16, we used the same 5 best-matching satellites in place of the corresponding MJ16 satellites, this time rotating the satellites by 90$^{\rm{o}}$ to obtain a control set of simulations to check the final distribution of the satellite debris. 

Each satellite contains hundreds of thousands DM particles and thousands of star particles. This, considering the number of satellites per simulation, allows for a good statistical investigation of the properties of the DM and stellar debris.   

We made use of the N-body code \texttt{GADGET-4} (Springel et al. 2020, in prep.). 
\texttt{GADGET-4} is a tree-code \citep{BarnesHut86} with additional SPH \citep{SpringelHernquist02,Hopkins13} and SF \citep{SpringelHernquist03} recipes; it is parallelized following Message Passing Interface (MPI) prescriptions. 
 
Typically, gas is stripped from satellites as they enter the MW environment \citep{Grebel03,Frings17,Simpson18}. Additionally, considering the median case, surviving gas in satellites forms the majority of their present-day stars before $z=1$ \citep{Weisz14}. 
Therefore, 
as a first approximation, we can switch-off the hydrodynamics and the SF recipes and focus on running N-body simulations only.

The M17 satellites are made of DM, stars and gas. In this work we were only interested in the tidal debris of the DM and the stellar component. Since the gas component contributes to the depth of the satellite potential wells, we kept it in the simulations as N-body particles. On the other hand the gas particles represent only a tiny fraction at each specific energy compared to the DM component (see Figure \ref{fig:sat3MassesDensprofEnergy}), which results also in a negligible contribution to the total debris. Therefore we added for simplicity the gas particles to the DM component in the following analysis.

In all our simulations we employed a softening $\epsilon = 200$ pc for the MW halo (in order to minimise spurious scatter effects on the DM halo particles) and $\epsilon = 50$ pc for the disc and bulge for a higher resolution in the force calculation. The satellite DM and star particles have softenings $\epsilon = 25$ pc and $\epsilon = 10$ pc, respectively. Given that for the employed gravitational softening kernel \citep{MonaghanLattanzio85} the Newtonian force radial dependence $\propto r^{-2}$ is exactly reproduced at distances $r > 2.8\epsilon$ \citep{Springel05}, the smaller softening choice adopted for the satellite particles allows for a very high resolution of the forces, and hence for a more accurate description of the tidal forces acting on the satellites.
The MW disc, bulge and halo particles have masses of $3.4\times 10^3$ M$_{\odot}$, $3.8\times 10^4$ M$_{\odot}$ and $4.4 \times 10^5$ M$_{\odot}$, respectively.
MJ16 also chose to use 10M particles in the disc of their MW models in order to have a high-resolution disc. This is useful because the higher resolution of the disc allows for less spurious scattering of the satellite debris particles that approach the inner MW halo, dominated by the disc. 

Each simulation is run on 96 parallel CPUs on the computer cluster \texttt{bwForCluster}. We run each of the 6 simulations for a total of 2 Gyr. For our analysis, we focused mainly on the final snapshot of each simulation. However, we generated outputs every 50 Myr, in order to track the process of mass stripping from the satellites.

\subsection{Impact of satellites on MW disc thickening and heating}

As a cross-check of the quality of our selection of satellites, we first compare the impact of our satellites on the MW disc with the corresponding results from MJ16 data.
MJ16, addressing the dynamical impact of the MW satellites on the Galactic disc, focused on the vertical thickening of the disc and on its heating (i.e., on the increase in time of the vertical velocity dispersion of its stars). 
We compare these results to check if a different distribution of satellites (the distribution of hybrid satellites) has a different impact on the thickening and heating of the disc, and hence if it produces a different dynamical effect on the disc. 

For each simulation we calculated the disc vertical thickness $z_{\rm{rms}}$ and the squared vertical velocity dispersion $\sigma_{z}^2$ for each ring-like bin of radius $R$ in the disc.
The vertical thickening and heating were calculated as 
$\Delta z_{\rm{rms}} = z_{\rm{rms},2} - z_{\rm{rms},0} $ and $
    \Delta \sigma_{z}^{2} = \sigma_{z,2}^2 - \sigma_{z,0}^2$
where the subscripts 2 and 0 are indicating that the given quantity is calculated at time $t=2$ Gyr and $t=0$ Gyr, respectively.

\begin{figure}[thb]
  \includegraphics[width=\linewidth]{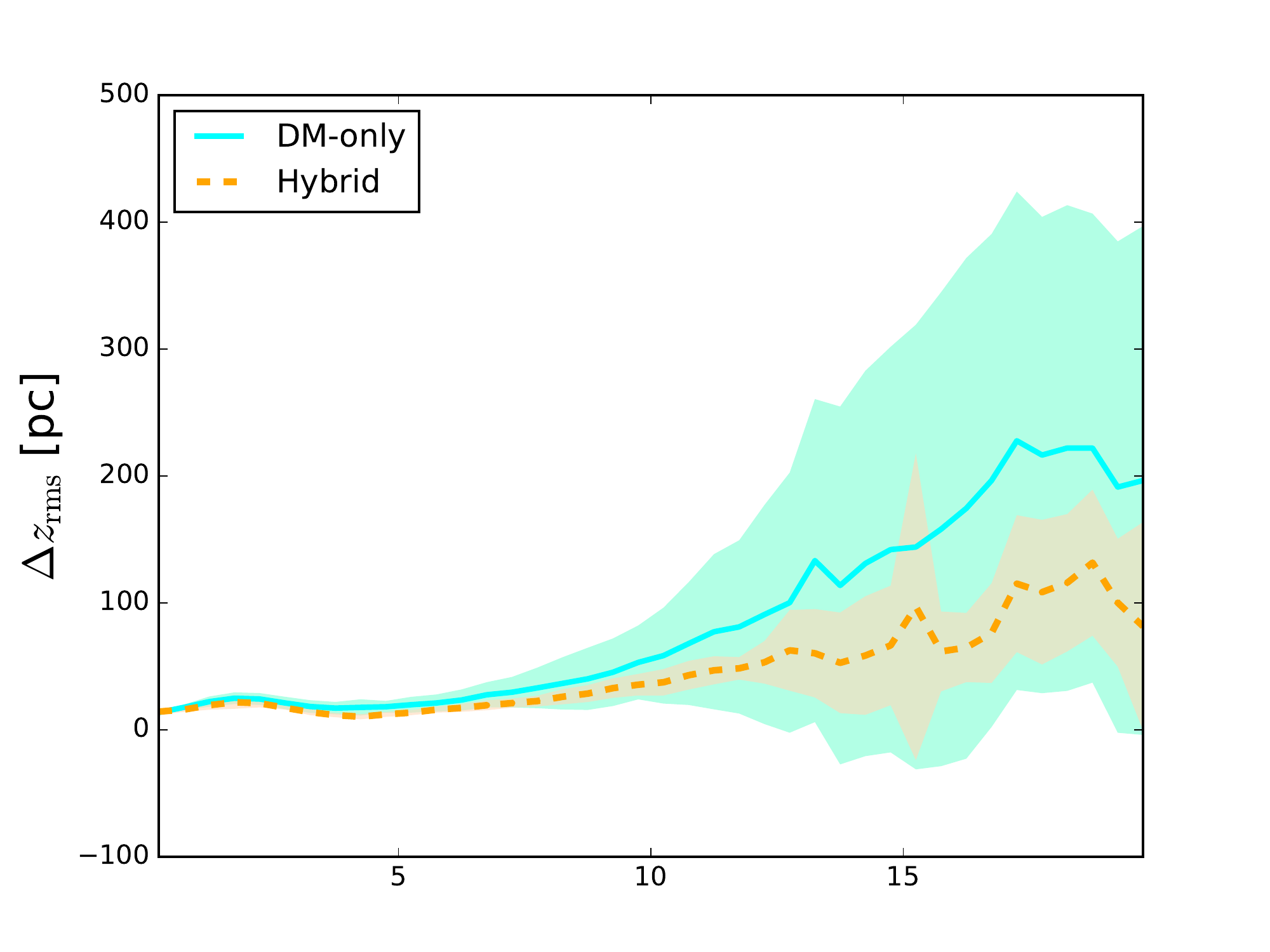}\\
  \includegraphics[width=\linewidth]{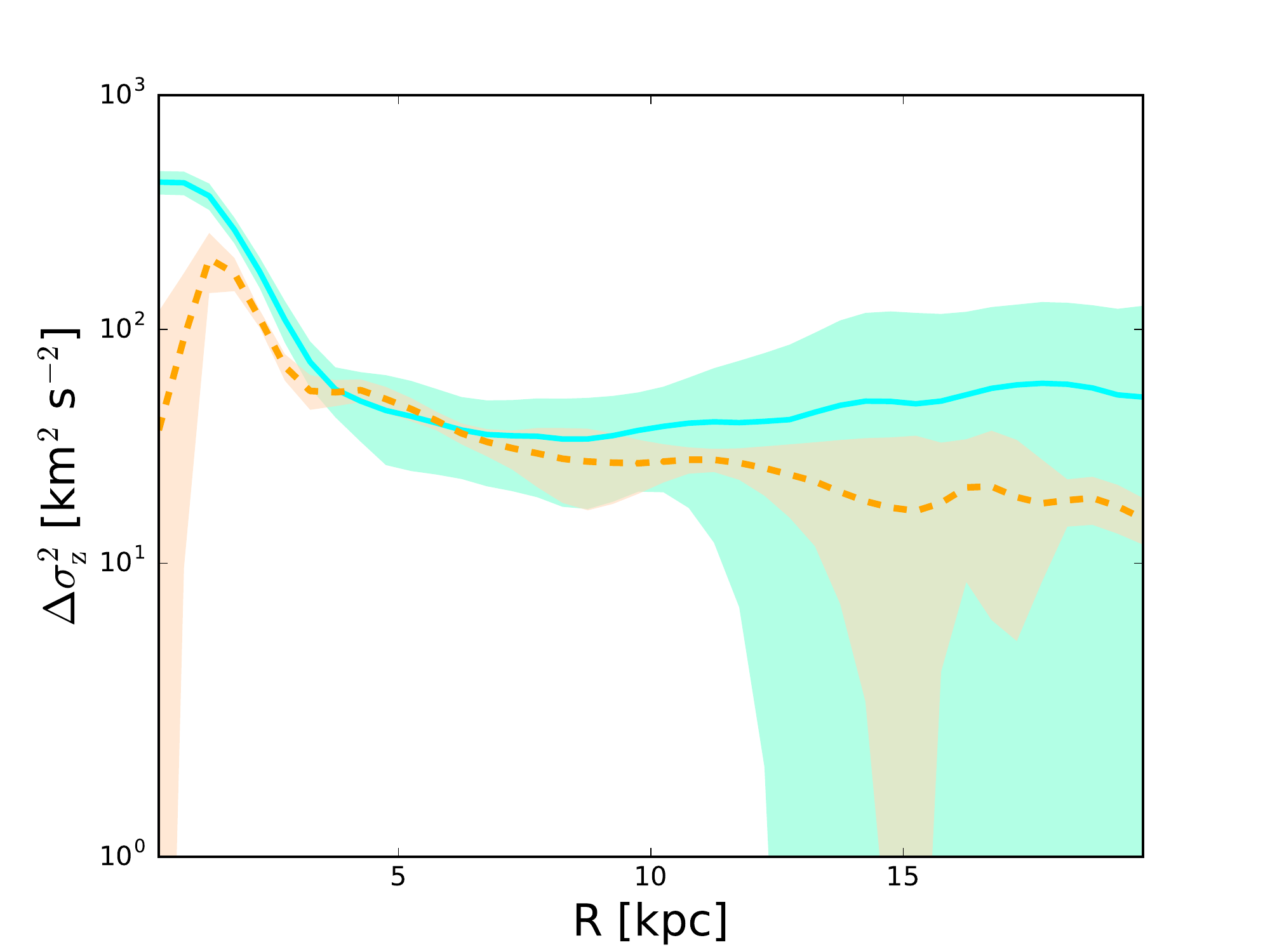}
  \caption{The top and the bottom panel show the radial profiles of disc thickening and disc heating, respectively. The cyan thick line represents the data from MJ16 simulations, the orange dashed line represents the data from our simulations.}
  \label{fig:impact_on_disc}
\end{figure}

In Figure \ref{fig:impact_on_disc} we show the radial profile of the disc vertical thickening and heating for the two simulation sets. 
We notice that, even if the disc thickening and heating are higher on average for MJ16 data, within the rms scatter between the six simulations our results are consistent with those from MJ16. 

The implications are that the MW disc is similarly excited if a population of DM-only satellites or hybrid satellites is utilised. 
Since the selection of satellites from MJ16 was done based on their dynamical effect on the MW disc, i.e., based on a mass cut, having a population of satellites with similar mass range (see Figure \ref{fig:1}) implies that the effects on the disc are similar and weakly dependent on the presence of DM only or of additional stars as well. The thicker scatter in MJ16 data is due to the F2 case, where a massive satellite of more than $10^{10}$ M$_{\odot}$ is strongly interacting with the MW disc.

\section{Results}

\subsection{Stripping of matter}

In this Subsection we show how much stellar and DM debris is stripped by the MW and their distribution in the Galactic environment.

\begin{figure}[thb]
  \includegraphics[width=\linewidth]{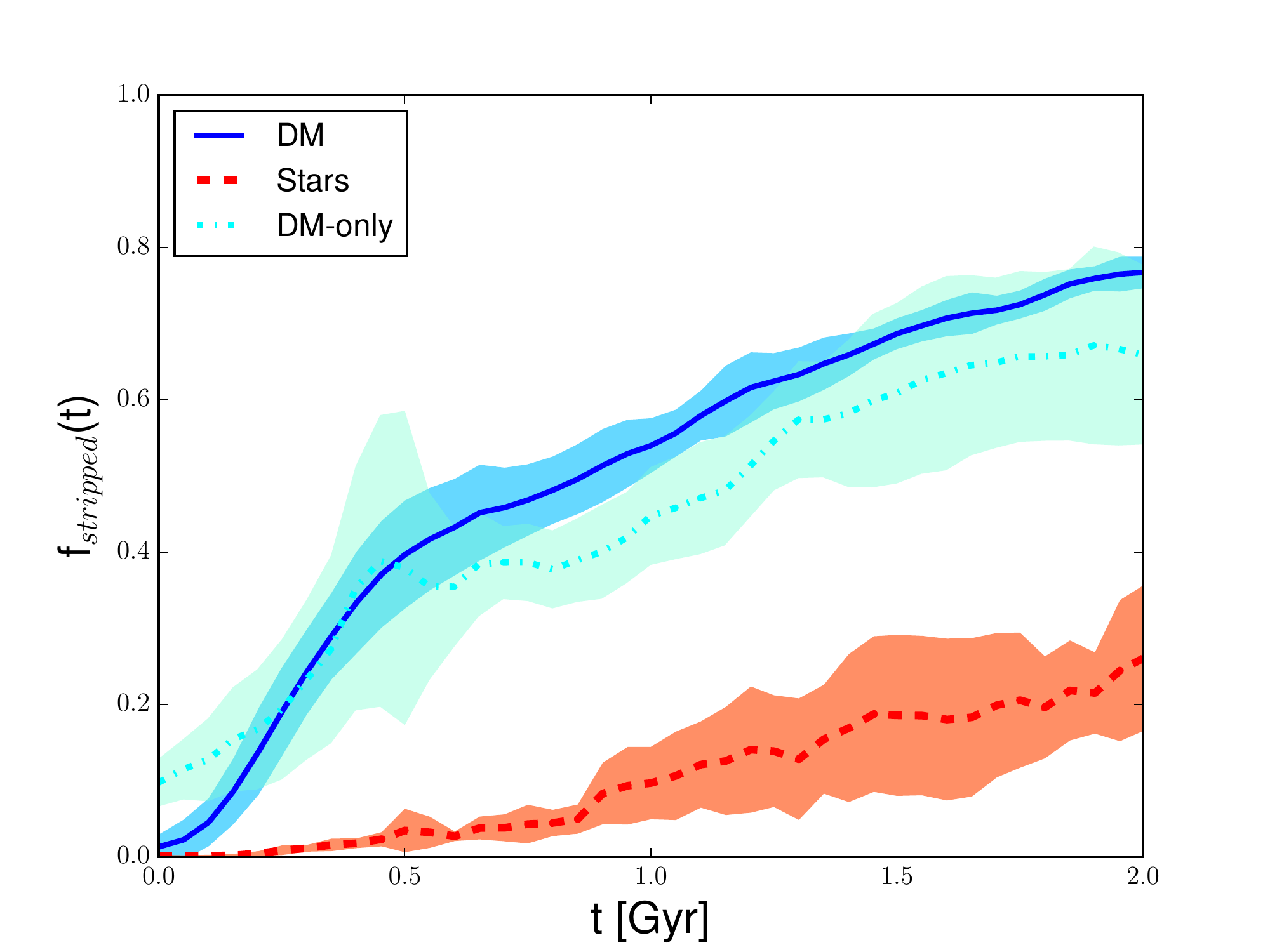}
  \caption{Fraction of the satellite matter stripped out of the tidal radii of all the satellites as a function of time. Colour codes and lines are dashed red for stars, thick blue for DM and dot-dashed cyan for the DM-only satellites from the MJ16 simulations. For each colour, the lines are the average between the six simulations, and the shaded areas represent the rms scatter.}
  \label{fig:tid_fraction}
\end{figure}

For each satellite, we calculated its density centre at every snapshot in order to define its position in time.
We used the tidal radius calculation of Equation \ref{eq:tidal_rad_definition} and at each snapshot we checked which fraction of total satellites stars and DM was found outside of the satellites tidal radii.
We show the result in Figure \ref{fig:tid_fraction}. The fraction of tidally stripped stellar debris increases to a maximum of 30$\%$ of the total, initial stellar satellite mass. Instead, the satellite DM is stripped up to $80\%$ of the total, initial mass. We also plot the results of our analysis on the data from MJ16. DM-only satellites also lose most of their DM, up to $70\%$ of the initial DM mass. 

In order to interpret these results, we focus on the properties and orbital distribution of the satellites in our simulations. These satellites were modeled as N-body objects made of star particles, mostly found in their core region, and DM particles, that have a shallower density profile and distribute up to large distances from the satellites density centres. Thus, DM can be stripped more easily and larger fractions of its mass, originally residing in the satellites, can end up as debris in the MW environment. Instead, stars are mostly confined in the inner regions of the satellites and significant stripping of this matter only occurs when the satellites closely approach the MW. 

\subsection{Radial distribution of the debris}

\begin{figure}[thb]
  \includegraphics[width=\linewidth]{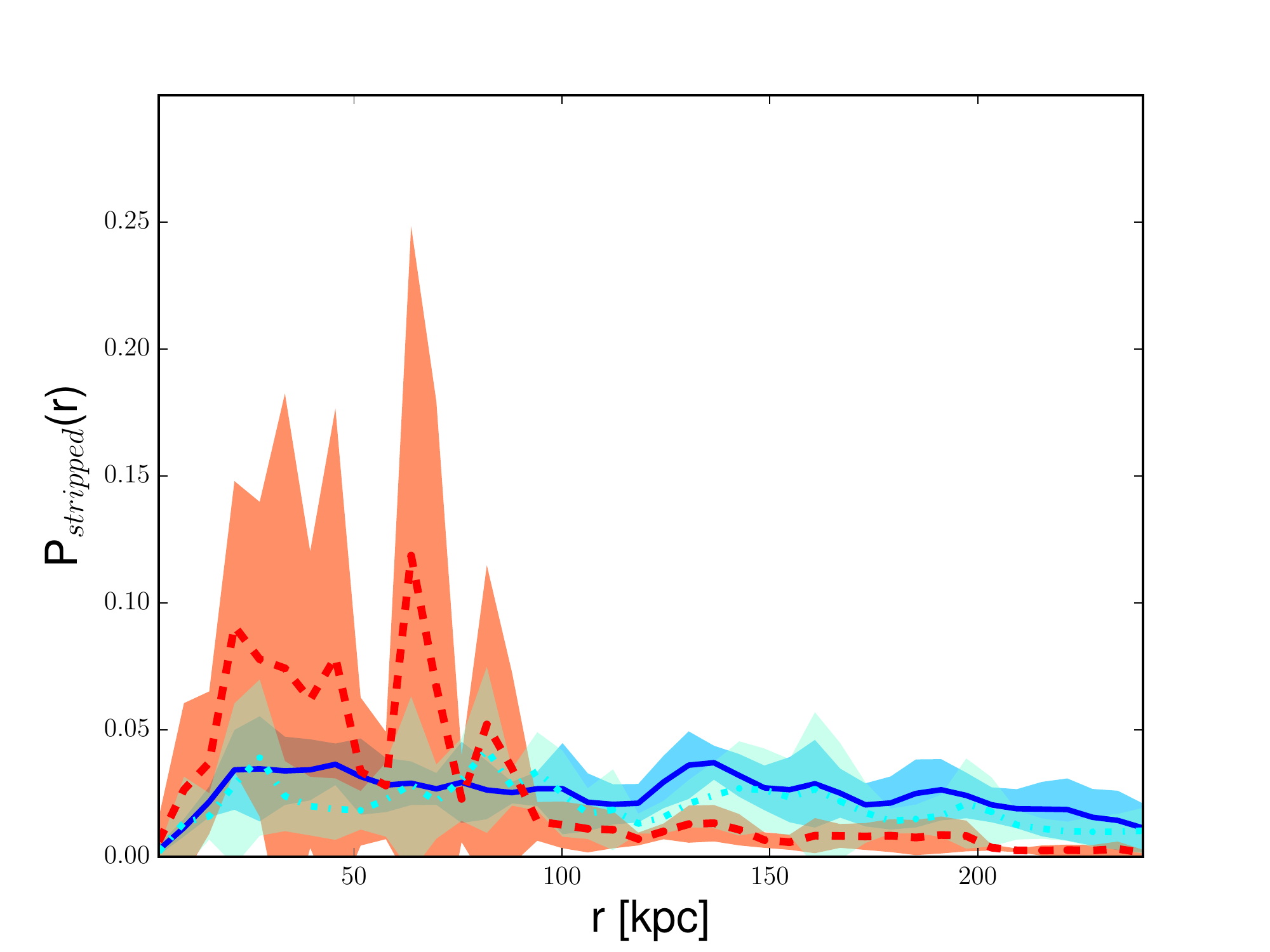}\\
  \includegraphics[width=\linewidth]{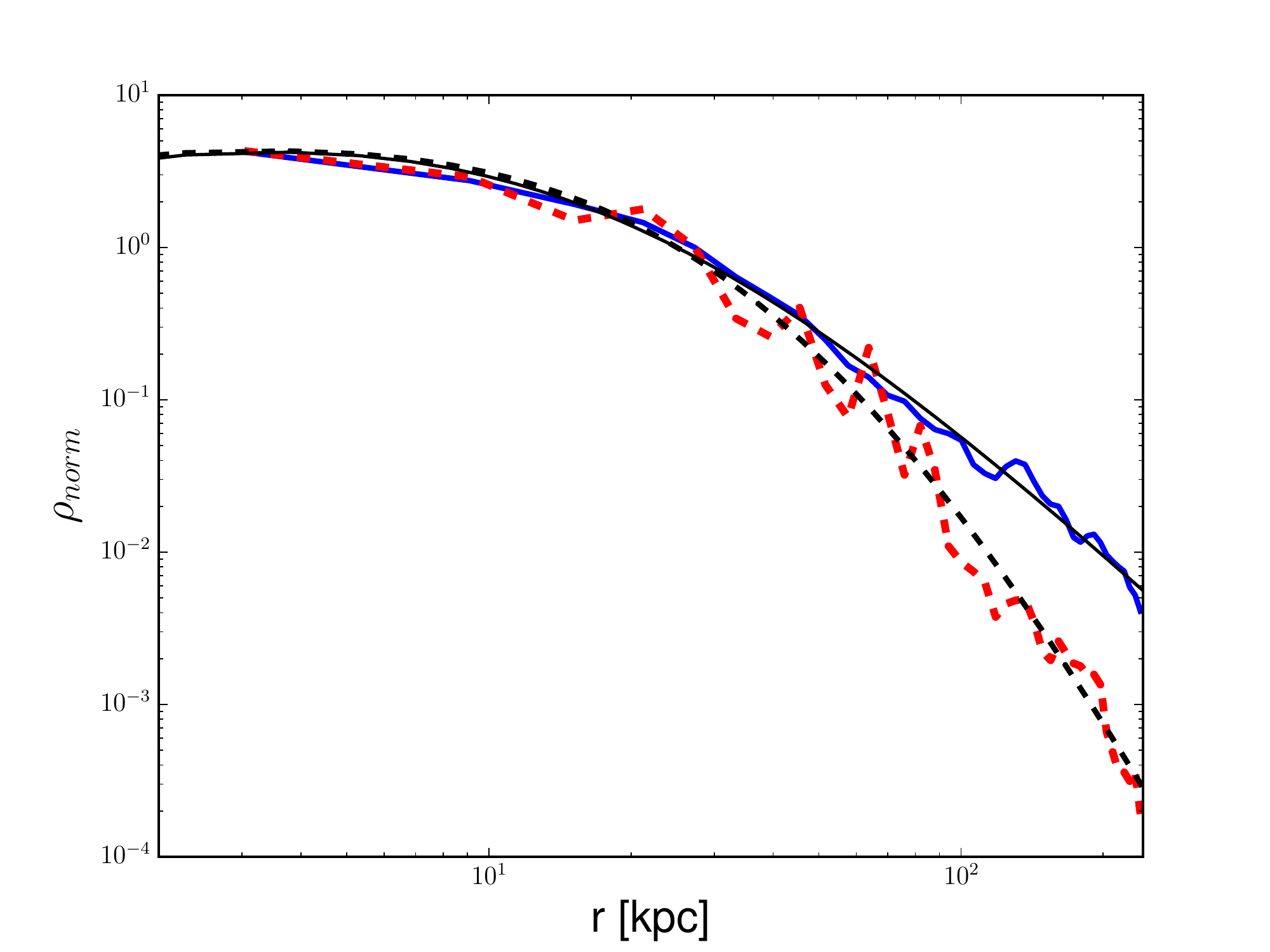} \caption{\textit{Top panel}: radial profile of $P_{\rm{stripped}}$ for the stripped debris mass deposited at a given spherical shell, normalised to the total debris mass of stars and DM, respectively.} Lines and colour codes are as in Figure \ref{fig:tid_fraction}. \textit{Bottom panel}: radial matter density profile of the debris in log-scale. Lines and colour codes are as in Figure \ref{fig:tid_fraction}. The DM-only satellites data from MJ16 are not plotted here. The thick black line and the dashed black line are the best fit profiles for the DM debris and stellar debris, respectively. All profiles are normalised to the corresponding density values at $r\sim$25 kpc.
  \label{fig:rad_fraction}
\end{figure}

Is there any difference in the final distribution of the stellar and DM debris in the MW environment? We addressed this by calculating the Probability Distribution Function $P_{\rm{stripped}}$ of the debris (normalised for the total debris mass within the virial radius of the MW; all the debris out of this radius is considered lost) that ends at a given GCd from the MW centre, as a function of the GCd. The radial bins have a width of $\sim 6$ kpc. $P_{\rm{stripped}}$ is a measure of how much mass out of the total debris ends in a given spherical shell.
In the top panel of Figure \ref{fig:rad_fraction} we plot the results of this calculation. 
The distribution of the DM debris is quite smooth and uniform throughout all radii up to the MW virial radius with some deficit in the innermost region. In the stellar distribution there are some prominent peaks in the inner 100 kpc which correspond to more confined stellar streams of satellites close to pericentre passage. In the outer MW halo, the stellar debris is found with significantly lower fractions. For comparison, the MW halo scale lengths in these MW models range between 20 kpc and 30 kpc (see Table \ref{tab:setup_MW}).
We note that DM-only satellites debris and hybrid satellites DM debris suffer similar tidal stripping and have similar values of $P_{\rm{stripped}}$. 

In the bottom panel of Figure \ref{fig:rad_fraction} we show the radial density profiles of the DM and stellar debris in the MW halo, as well as their best fitting functions, for the simulations with the hybrid satellites. The radial density profiles were calculated dividing the values of $P_{\rm{stripped}}$ by the spherical-shell volumes of the corresponding bins. To fit the data, we chose the generalised NFW profile as a function of GCd
\begin{equation}
    \rho(r) = \frac{\rho_0}{\bigg(\frac{r}{r_{\mathrm{s}}}\bigg)^{\alpha} \bigg( \frac{r}{r_{\mathrm{s}}} + 1 \bigg)^{\beta-\alpha}} \,\, .
\end{equation}
Here, $\rho_0$ is the scale density, $r_{\mathrm{s}}$ is the scale radius, $\alpha$ controls the inner slope and $\beta$ controls the outer slope. 
The best-fit parameters are given in Table \ref{tab:bestFit}. For both debris components we found that the inner slope is positive ($\alpha < 0$) describing the mass deficit inside $\sim$5 kpc for DM and for stars.
The profile scale radius is higher for stars, reaching $\sim45$ kpc against the $\sim15$ kpc of the DM debris profile. The outer slope of the DM debris is very close to that of a standard NFW profile. In contrast, the stellar debris shows a much steeper drop at large radii.

This picture points to the fact that at large radii less stars are stripped and the stars stripped in the inner halo are on more circular orbits. 
\begin{table}[t!]
\caption{\footnotesize Best-fit parameters for the fitting curves of the density profiles of DM and stellar debris in Figure \ref{fig:rad_fraction}.}  
\centering      
\begin{tabular}{c | c c c c}          
\hline\hline                        
 \\
Debris & $\rho_0$  & $r_{\rm{s}}$ & $\alpha$ & $\beta$ \\
component & (M$_{\odot}$ kpc$^{-3}$) & (kpc) & & \\
\hline                          
    DM & $1.62 \times 10^5$ & 15.34 & -0.66 & 2.94\\ 
    Stars & $7.15 \times 10^1$ & 47.51 & -0.44 & 6.17\\ 
\hline 
\end{tabular}
\label{tab:bestFit}  
\end{table}

This also explains the inner peaks of stellar mass fractions in Figure \ref{fig:rad_fraction}. DM debris can be released at any distances and the contribution to its radial distribution comes from both inner and outer satellites, thus the Probability Distribution Function of the DM debris is radially more uniform and its density profile is less steep.  
Calculating the cumulative fraction of the stellar debris as a function of GCd, we find that $\sim30\%$ of the total stellar debris is inside 30\,kpc and $\sim50\%$ is within 60\,kpc of GCd. 

\subsection{Shape of the debris}

Now, we ask what the geometrical distribution of the tidal debris is in the inner MW region. We focused on the inner 25 kpc of GCd, in order to see what the impact of the local MW environment is on the debris. We wanted to understand the following points: \textit{a)} does the debris finally show a spherical geometry, or a flat one? \textit{b)} What is the orientation of the debris distribution? \textit{c)} Is there any difference between the DM and stellar debris? 

In order to answer these questions, we introduced the Second Order Momenta Tensor (SOMT hereafter) in our analysis. The SOMT is a rank-2 tensor for which the $jk$ entry is calculated as

\begin{equation}
    \begin{split}
        I_{jk} = \sum_{i=1}^{N}{m_i x_{i,j} x_{i,k}} \, ,
    \end{split}
\end{equation}
where $i$ is indicating the $i-$th particle (out of $N$ particles), $m_i$ is its mass and $x_{i,j}$, $x_{i,k}$ are its cartesian coordinates \citep{BinneyTremaine08,Polyachenko16}. In our case $N$ is the total number of particles that fall in a given sphere centered on the MW Galactic centre.
The definition employed here for the SOMT is addressing the global geometrical distribution of the debris, no matter how sub-structured it is.

An indicator of the flattening of the matter distribution is represented by the ratio $c/a$ between the SOMT semi-minor and semi-major axes $c$ and $a$, whereby $c$ and $a$ are the square roots of the eigenvalue with the smaller and larger magnitude, respectively.

\begin{figure}[thb]
  \includegraphics[width=\linewidth]{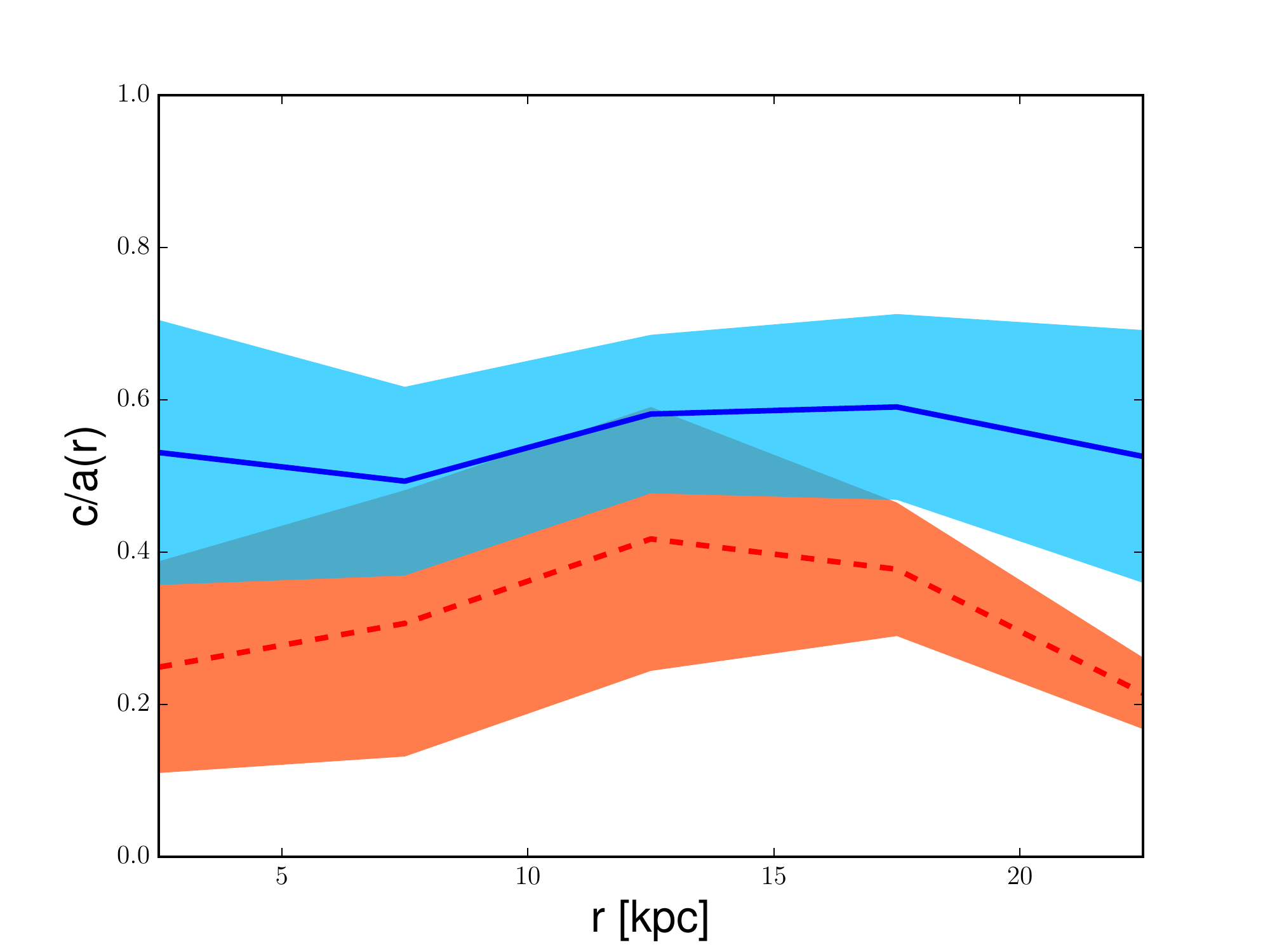}
  \caption{Radial profile of the $c/a$ ratio for the SOMT of the stellar (red) and DM (blue) debris after 2\,Gyr simulation. Lines and colour codes are as in Figure \ref{fig:tid_fraction}. For each profile, the central line is the average profile between the 6 simulations. The shaded areas represent the rms scatter. }
  \label{fig:ca_ratio}
\end{figure}

We focused on the inner 25 kpc in the MW halo, since we were interested in the behaviour of the debris in the local halo environment.
Specifically, for each simulation we calculated the tensor for the debris falling in each sphere centered on the Galactic Centre. We chose a radial binning of 5 kpc to minimise the noise due to low number statistics and to smooth the contribution of single streams. 
For each simulation we calculated $c/a$ at each sphere and then for each sphere we averaged the ratio among the six simulations. 
In Figure \ref{fig:ca_ratio} we show the radial profile of $c/a$ for the stellar and DM debris together with the root mean square (rms) scatter. 
From Figure \ref{fig:ca_ratio} it is evident that both the stellar and DM debris are geometrically flat but with a significant scatter. For the DM debris we found $c/a\sim 0.55$, whereas the stellar debris is flatter by a factor of two. The flattening depends only weakly on the size of the sphere. 

\begin{figure}[thb]
    \includegraphics[width=\linewidth]{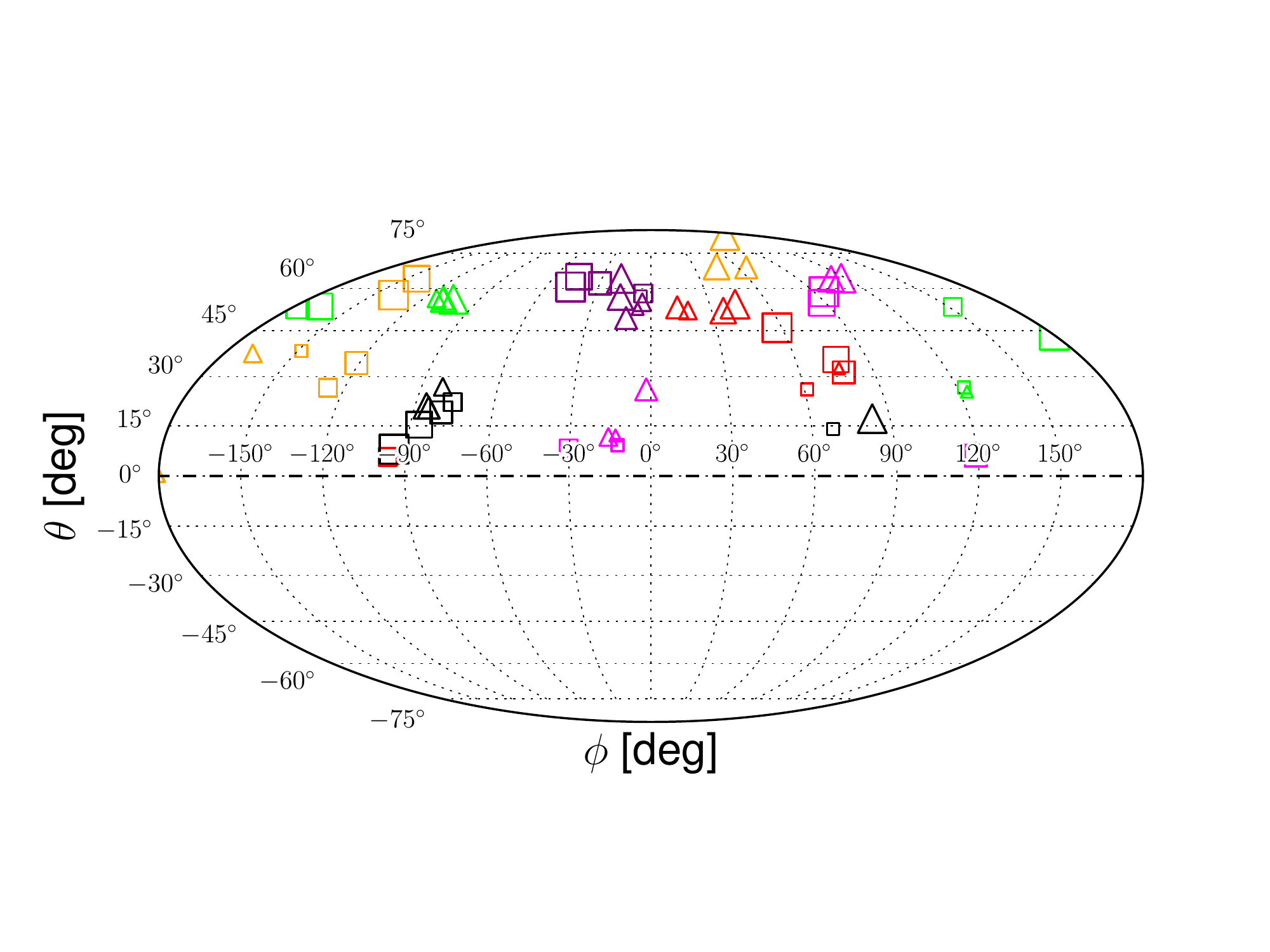}
  \caption{Mollweide projection of the distribution of the orientation of the semi-minor axis $c$. The azimuth $\phi$ changes horizontally, while the latitude $\theta$ changes vertically. Squares represent DM, triangles represent stars. Smaller markers are for inner radii, larger markers are for outer radii. Colours are as in Figure \ref{fig:1} and correspond to the simulations with different initial conditions. The horizontal dot-dashed line indicates the Galactic plane.}
  \label{fig:c_mollweide_all}
\end{figure}

Now, we ask what the spatial orientation of the flattening is, which is defined by the direction of the shortest eigenvector $e_c$ of the SOMT corresponding to the eigenvalue $c^2$. We calculated this orientation in the $\theta-\phi$ (latitude-azimuth) angles-space for the DM and stellar debris at each radius and for each simulation separately.
In Figure \ref{fig:c_mollweide_all} we plot
 the distribution of the directions of $e_c$ in the angles-space at different GCds, using a Mollweide projection.
If $e_c$ was defined at any point with negative $z$ corresponding to $\theta<0$, we changed its sign in order to have it with positive values of $z$.
This was done to avoid ambiguities in the interpretation of the direction of the minor axis in the angles-space, since every eigenvector allows to take both orientations along its direction, with a difficult interpretation in the Mollweide projection\footnote{In fact, the opposite of a vector in this kind of projection is not simply represented as the symmetric opposite in the map with respect to $(\phi,\theta)=(0,0)$.}. 

The orientations of the short axis in the different simulations (different colours) are distributed over a wide region in the angles-space. 
This suggests that there is not a huge impact of the disc on the final distribution of the debris. 
Additionally there is a large variation of the orientation with increasing GCd (size of symbols) as a sign of strong substructures in the debris.
Furthermore, squares (DM) and triangles (stars) of the same simulation (same colour) do not occupy the same region in the angles-space. We interpret this as an indication that the DM and stellar debris show significantly different distributions and structures. 

\begin{figure}[thb]
  \includegraphics[width=\linewidth]{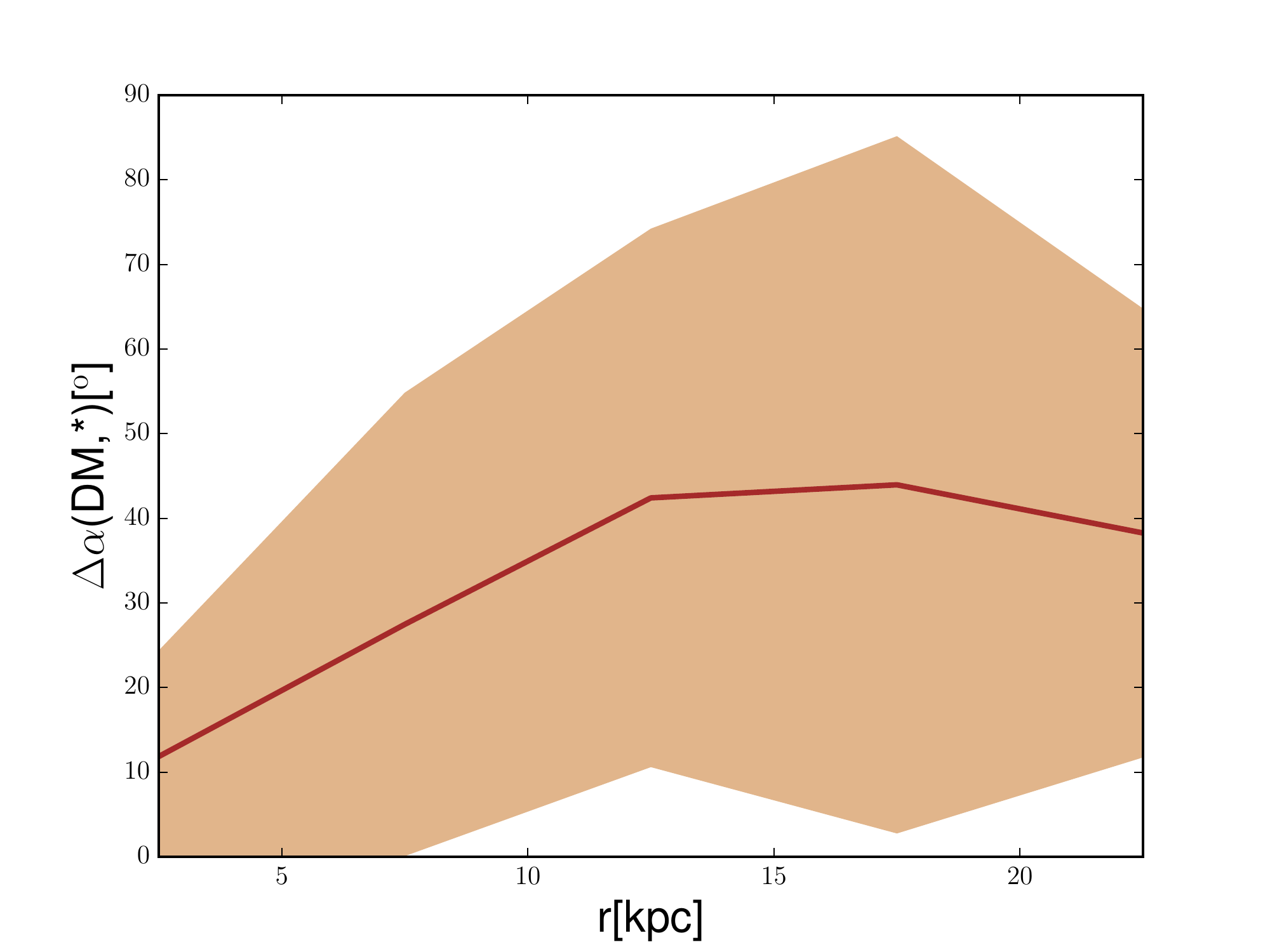}
  \caption{Average radial profile of the GCircD between DM and stellar debris as a function of spherical radius $r$. The shaded area represents the rms scatter.}
  \label{fig:difference_in_angle_DM_star}
\end{figure}

We wanted to quantify how big the differences in $\phi$ and $\theta$ are between the short axis of the stellar and DM debris tensor. To do this, we calculated the Great-Circle Distance (GCircD) $\Delta \alpha$ at each sphere between DM and stars by
\begin{equation}
\label{eq:angle_distances-GreatCircle}
    \Delta \alpha(\mathrm{DM},*) = \arccos{\Big[ \sin {\theta} _{\mathrm{DM}} \sin{\theta}_* + \cos\theta_{\mathrm{DM}} \cos{\theta}_*\cos \Delta \phi \Big]}\,
\end{equation}
where $\phi_{\rm{DM}},\theta_{\rm{DM}}$ are the angles of $e_c$ for the DM debris SOMT as a function of the radius $R$ of the sphere; $\phi_*,\theta_*$ are the same angular quantities for the stellar debris; and where $\Delta \phi = \phi_{\rm{DM}} - \phi_*$. 
We show these results in Figure \ref{fig:difference_in_angle_DM_star}.
The average difference in angular distribution between DM and stars is no more than 10-20$^{\rm{o}}$ in the central part of the halo (first few kpc), whereas it reaches up to 40$^{\rm{o}}$ going out towards 15-20 kpc, where the scatter around the average is very high. 
This indicates that it is not possible to find a simple, systematic correlation between the DM and stellar debris orientations in the environment around the MW. This is reinforced by the different $c/a$ and the different radial distribution of the debris seen in Figures \ref{fig:rad_fraction} and \ref{fig:ca_ratio}. The DM and stellar debris do not share the same geometry once they are stripped from their satellite progenitors. Observationally, this means that it is not possible to track the distribution of the DM debris from the distribution of the stellar debris directly. 

It is important at this point to understand if the disc can have any impact at all on the final distribution of the debris, or if this one depends on the satellites initial conditions only.
To test this, we ran another set of six simulations, each corresponding to one of the six previous simulations. For each simulation setup, we rotated the initial relative position and velocity of the disc particles by 90$^{\rm{o}}$ around the $y$-axis resulting in a disc in the $y-z$ plane rotating around the x-axis\footnote{ This effective rotation transforms the $x$-axis to the $-z$-axis, the $z$-axis to $x$-axis and that leaves the $y$-axis unchanged.}.
We ran these new six simulations for 2 Gyr. 

If the disc were mainly responsible for the flattening of the satellite debris, we would expect the minor axis distribution of the SOMT to be rotated approximately similar to the disc by $\sim$90$^{\rm{o}}$.

\begin{figure}[thb]
\includegraphics[width=\linewidth]{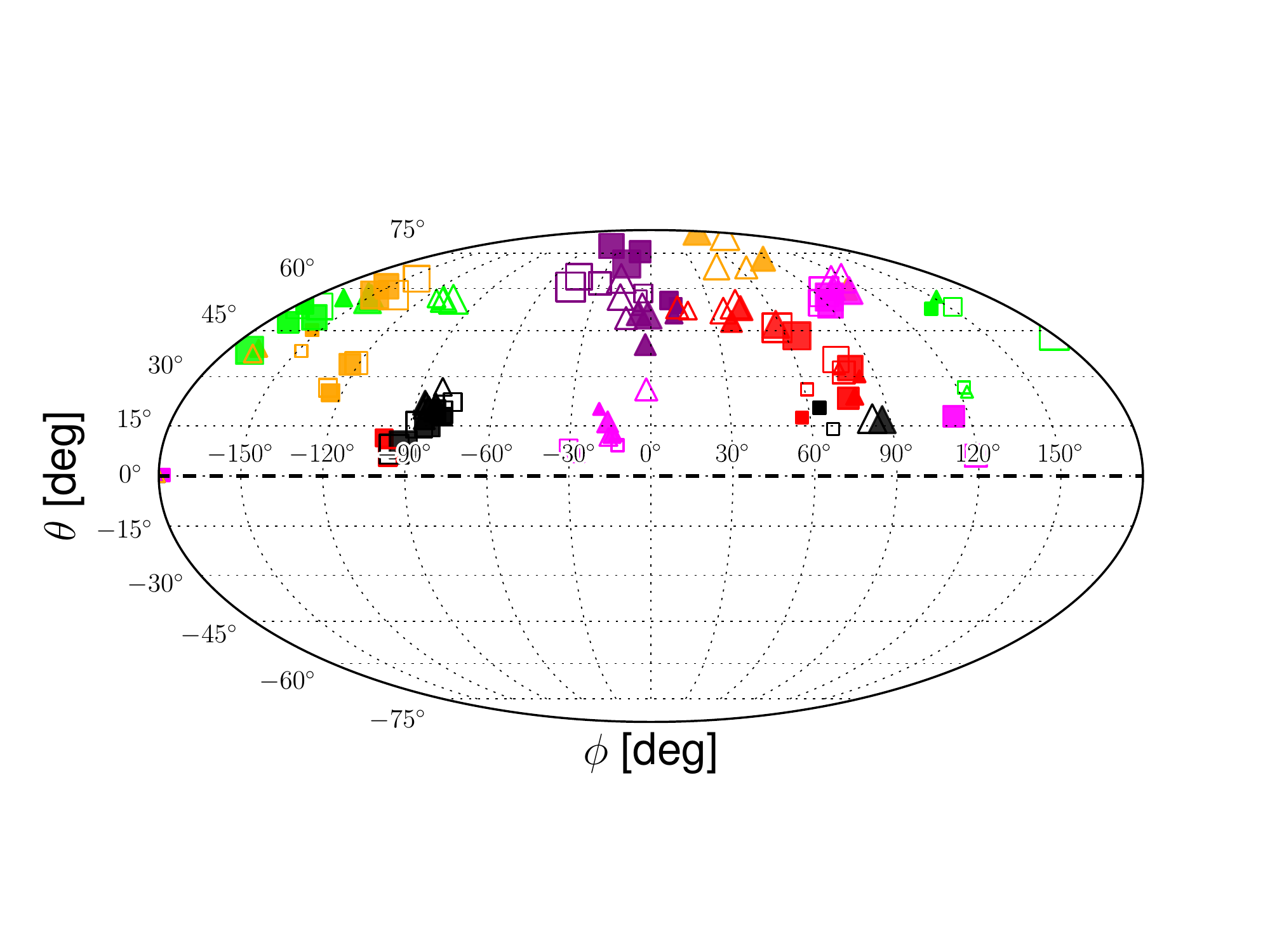}
\caption{Mollweide projection of the distribution of the $c$ axis in the angles-space, for the rotated and non-rotated disc. Data from Figure \ref{fig:c_mollweide_all} are re-plotted as empty markers. Solid markers are for rotated disc simulations.}
  \label{fig:ca_ratio_aver}
\end{figure}

At the end of the simulation, we analysed the final angular distribution of the debris. As can be seen in Figure \ref{fig:ca_ratio_aver}, the final Mollweide projection of the debris is mostly overlapping the same region of the original debris distribution showing no systematic rotation. This confirms that the initial conditions of the satellites have a larger effect than the disc orientation on the final debris distribution. 

We then quantified the GCircD at each sphere between the debris in the new set of simulations and the debris from the original set of simulations
\begin{equation}
\label{eq:angle_distances-GreatCircle_2}
    \Delta \alpha(\mathrm{rot},0) = \arccos{\Big[ \sin\theta_{\mathrm{rot}} \sin\theta_0 + \cos\theta_{\mathrm{rot}}\cos\theta_0\cos\Delta \phi\Big]}
\end{equation}
where in this case $\phi_{\mathrm{rot}},\theta_{\mathrm{rot}}$ are the angles of the $c$ minor axis for the SOMT within the given spherical radius $r$ for the debris from the second set of simulations with rotated disc orientation; $\phi_0,\theta_0$ are the same angular quantities for the original set of simulations; and where $\Delta \phi = \phi_{\mathrm{rot}} - \phi_0$. 

\begin{figure}[thb]
  \includegraphics[width=\linewidth]{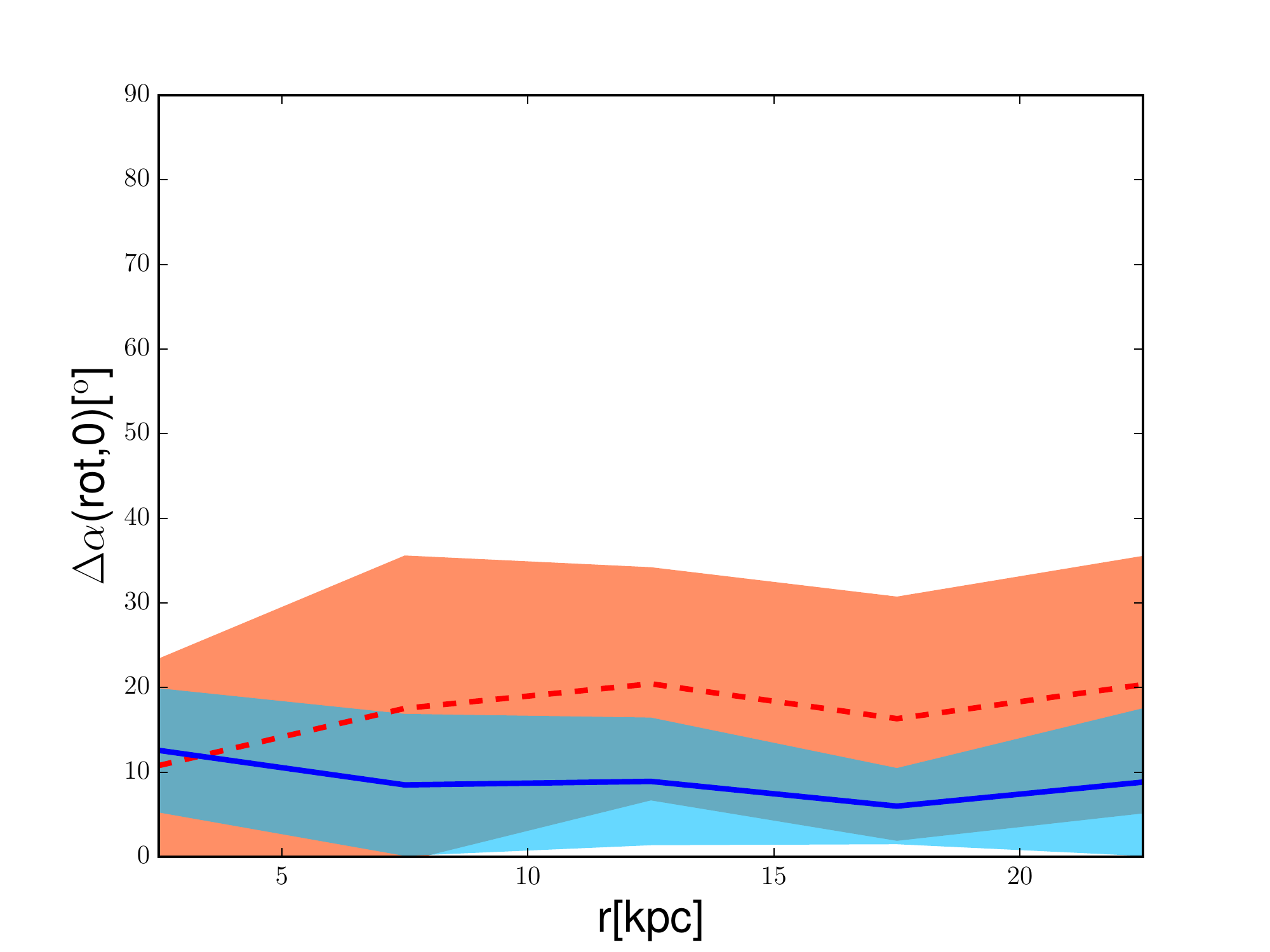}
\caption{Radial profile of the GCircD $\Delta \alpha$ of the rotated disc simulations compared to the corresponding, original simulations, for the short axis orientation of the stellar and DM debris. Lines and colour codes are as in Figure \ref{fig:tid_fraction}. The shaded areas are again the rms scatter.}
  \label{fig:plot_great-circle_average}
\end{figure}

In Figure \ref{fig:plot_great-circle_average} we can see that on average $\Delta\alpha(\mathrm{\mathrm{rot}},0) \sim 10^{\rm{o}}$ for DM, while at the outer radii it reaches values of up to $\Delta\alpha(\mathrm{rot},0) \sim 20^{\rm{o}}$ for stars. 

So, the rotation of the disc initial conditions produces a limited, yet not completely negligible rotation of the final distribution of the debris that overall implies that the disc may have a moderate impact on the final distribution of the local satellite debris.

\section{Additional investigation: fraction of surviving satellites and DM/stellar mass ratio}

\begin{figure}[thb]
  \includegraphics[width=\linewidth]{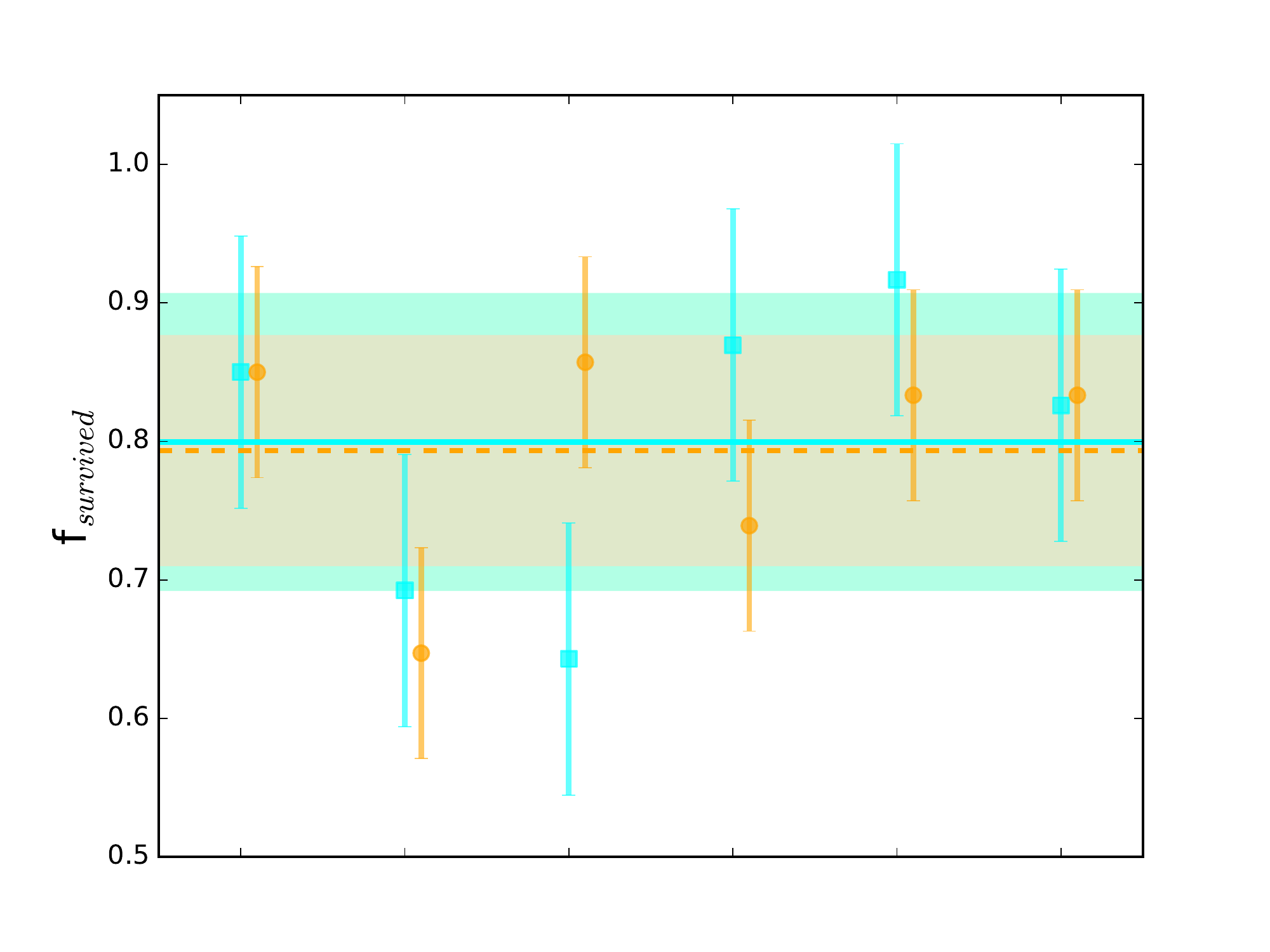}\\
  \includegraphics[width=\linewidth]{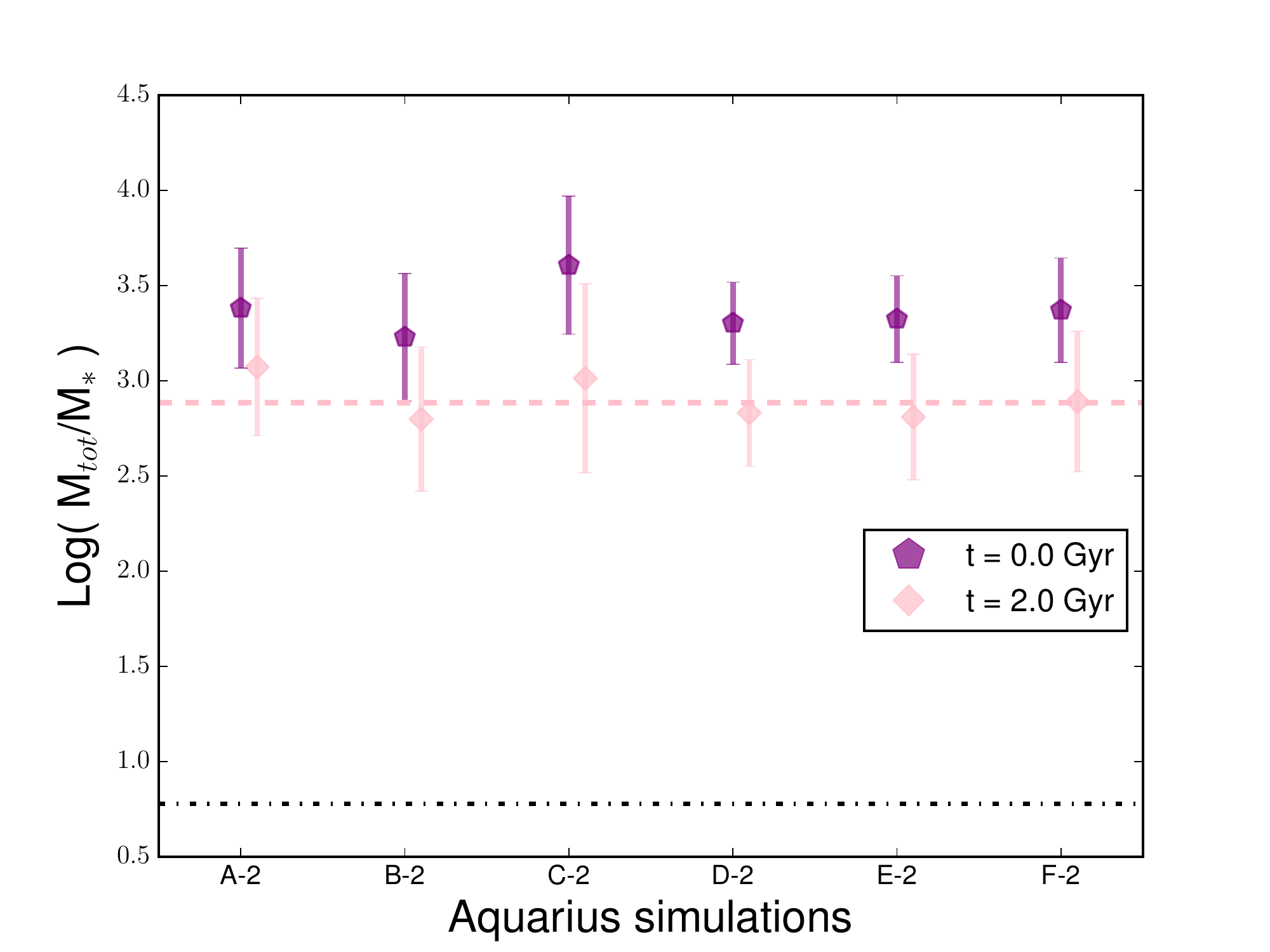}
  \caption{\textit{Top panel}: fraction of surviving satellites, with $10\%$ of their initial mass adopted as the threshold to determine their survival. 
  Colour codes are as in Figure \ref{fig:impact_on_disc}.
  The orange and cyan errorbars are the rms scatter for our set of simulations and for MJ16 simulations, respectively. 
  \textit{Bottom panel}: logarithm of the ratio between total mass and stellar mass in the satellites in Aq-A2 to F2 simulations, $\log{({M}_{tot}/{M}_{*})}$, for the initial and final snapshot. Purple pentagons are the average of each corresponding simulation for the initial data at $t=0\,\rm{Gyr}$, pink diamonds are instead for the final data at $t=2\,\rm{Gyr}$. The purple and pink errorbars are the rms scatter for all the satellites in each corresponding simulation (initial and final snapshot, respectively). The dot-dashed black line is the density ratio $\Omega_{\rm{m}}/\Omega_{\rm{bar}}$ between all matter and baryonic matter. The pink dashed line represents the best-fit for the final ratios, ($M_{\rm{tot}}$/$M_{*}$)$_{\rm{best}}$. 
  Only the satellites that survive the stripping process are employed to calculate the average and the standard deviation in each simulation.}
  \label{fig:frac_10perc}
\end{figure}

In addition to the final distribution of the debris in the MW halo, we checked the fraction of satellites that after 2 Gyr survived stripping from the MW. 
To determine if a satellite survived stripping or not, we chose a threshold fraction of $10\%$ of its total, initial mass. A satellite that has a final tidal mass larger than $10\%$ of its initial mass is considered to have survived. Otherwise, it is counted as dissolved. 
In Figure \ref{fig:frac_10perc}, top panel, we show the fractions of the survived satellites for our six simulations and for MJ16 simulations. The error-bar around each data is the scatter between the six simulations of the same set.
 
We found that in all simulations more than $\sim 65\%$ of the satellites survived after 2\,Gyr. The averages between the six simulations for both cases are reported as a thick cyan line (DM-only) and a dashed orange line (hybrid). We also plot the area of the rms scatter between them with the same colour code. We found $f_{\rm{survived}} \sim 0.8 \pm 0.1 $ for both sets of simulations, thus we confirm the similarity of results between them with no systematic differences.

We also looked at the ratio between DM and stellar mass inside the satellites at the beginning and at the end of the hybrid simulations, to understand the evolution of the matter content inside the satellites. This result is plotted in the bottom panel of Figure \ref{fig:frac_10perc} in logarithmic values. The best-fit value $(M_{\rm{tot}}/M_{*})_{\rm{best}} = 768 \substack{+217 \\ -301}$ at the final epoch is also shown.
For all the survived satellites the ratio between DM and stellar matter, though still much above the cosmic ratio, decreased in time. This is a consequence of what was shown in Figure \ref{fig:tid_fraction} where more DM than stars is stripped from the satellites.

For comparison, we give the cosmic ratio between all matter and baryons from Wilkinson Microwave Anisotropy Probe (WMAP)-7 results \citep[see][]{Komatsu11}, $\Omega_{\rm{m}}/\Omega_{\rm{bar}} \sim 6.0$ being much smaller than the initial ratio in our simulations. 
It appears that the DM-star ratio is subject to two phases. As argued in M17, in the first phase, before they strongly interact with the MW, the satellite galaxies lose large quantities of gas, therefore their total-matter-to-baryons ratio increases (so they start in our simulations with a high total-matter-to-stars ratio). In a second phase, i.e., in more recent epochs, strong tidal interactions with the MW deplete the satellite dwarf galaxies of more DM than stars and drive the total-matter-to-stars ratio towards lower values by a factor of 2--4 on average.

Assuming a stellar mass-to-light ratio of two in solar units, the satellites in our simulations fall in the regime $1.2 \times 10^3 \rm{L}_{\odot} - 4 \times 10^6 \rm{L}_{\odot} $ corresponding to UFDs \citep[hereafter S19]{Simon19}.\footnote{There is one satellite in our simulation from Aq-F2 setup that ends with L$= 50$ L$_{\odot}$, but it is below the range considered in Figure 4 of S19.}
\begin{figure}[thb]
  \includegraphics[width=\linewidth]{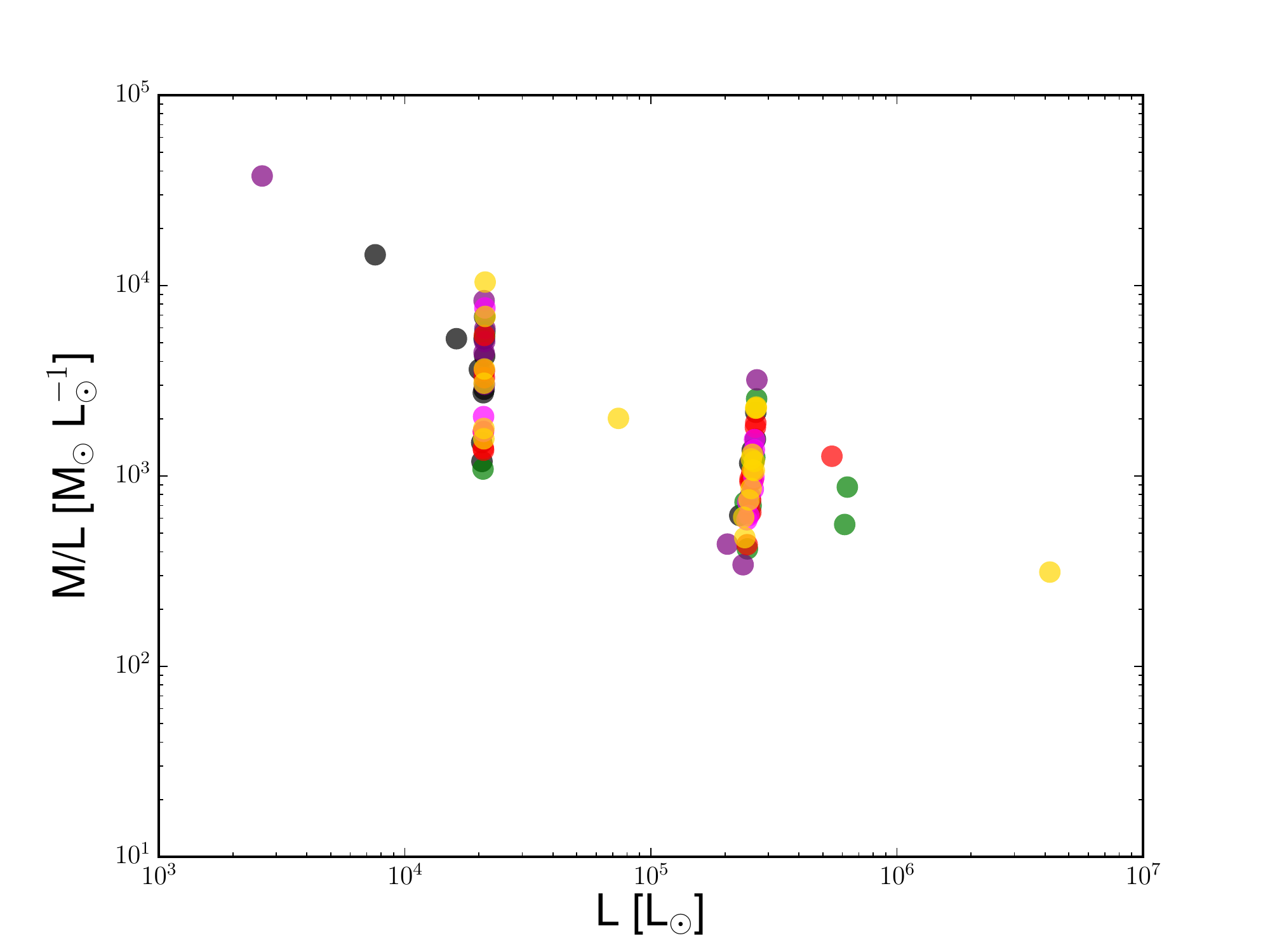}\\
  \caption{Distribution of the survived satellites in the $L-M/L$ plane for the six simulations adopting a stellar mass-to-light ratio of two. Colour codes are as in Figure \ref{fig:1}.}
  \label{fig:ML}
\end{figure}
The mass-to-light ratios, which correspond to the tidal mass-to-light ratios, of all our surviving satellites are shown as function of luminosity in Figure \ref{fig:ML}. 
The tidal mass-to-light ratio was estimated for the first time by \cite{FaberLin83} for 7 dwarf spheroidal galaxies, finding a much smaller value than our mass-to-light ratio of ($M/L$)$_{\rm{best}} \sim (1.5 \times 10^3) \substack{+434\\-606} $ M$_{\odot}$ L$_{\odot}^{-1}$.
S19 showed that $M/L$ inside the half-light radius $R_{\mathrm{h}}$ decreases with increasing luminosity from $\sim 1000$ (with a large scatter) for $L= 10^{3}$ L$_{\odot}$ to $\sim 10$ for $L= 10^{7}$ L$_{\odot}$, which is similar to the values of \citet{Errani18} inside $1.8\,R_{\mathrm{h}}$. We found the same trend of mass-to-light ratios as function of luminosity, but our values are systematically higher by a factor of a few. This offset is expected, because the tidal radius is larger than $2\, R_{\mathrm{h}}$ in most cases.

\section{Conclusions and discussion}

\subsection{Conclusions}

We have performed a set of N-body simulations to investigate the general properties of the satellite debris distribution in the MW environment. For the first time, we took advantage of a set of satellite galaxy models taken from cosmological high-resolution hydrodynamical simulations, placed in initial orbits derived from the results of N-body cosmological simulations of structure formation, and of a full high-resolution N-body MW host model taken from previous literature, consisting of live disc, bulge and halo. For a statistical analysis, the initial conditions of six Aquarius simulations were used. Following the approach of MJ16, all satellites with tidal masses larger than $10^8\,$M$_{\odot}$ and with pericentre passage closer than 25\,kpc were taken into account. We ran each simulation for 2 Gyr.

We investigated general properties of the debris of satellite galaxies in the global and local MW environment. We focused on the differences in the distribution of stellar and DM debris.
Based on our findings, we can state that:
\begin{itemize}
    \item the stellar component in the satellite galaxies is much more tightly bound than the DM component and cannot be deduced from DM-only simulations;
    \item the stripping process acting on satellites DM is more efficient than on satellite stars and releases more DM debris than stellar debris in the MW environment (80\% compared to 30\%);
    \item the radial density profile of the DM debris covers the whole host halo and shows a standard NFW slope in the outer region; the stellar debris is confined to the inner 50\,kpc with a steep cutoff outside; both profiles show a deficit in the inner 5\,kpc;
    \item the stellar debris shows more prominent peaks of the radial Probability Distribution Function than the DM debris, pointing to a more confined structure of individual streams in the inner part of the MW halo; 
    \item the debris of DM and stars distribute with some degree of flatness ($c/a\sim$ 0.55 and 0.3, respectively); the orientation of the minor axis is very different for the different simulations and shows no obvious correlation to the MW disc plane; 
    the orientation of the DM and the stellar debris is not well correlated in each simulation; thus 
    it is not possible to reconstruct DM debris and streams directly from observed stellar streams;
    \item changing the orientation of the disc by 90$^{\rm o}$ has a small effect on the distribution of the satellite debris; this confirms that the structure of the DM and stellar debris of satellite galaxies is mainly determined by the initial conditions of the satellites; the flattened potential of the disc plays a minor role only;
    \item the tidal total-to-stellar mass ratio of the satellites decreases by a factor of 2--4 during the simulations and shows at the end mass-to-light ratios, which are consistent with observations of the Local Group UFDs.
\end{itemize}

In conclusion, our work states that satellites stars and DM are subject to different stripping efficiency and different redistribution in the MW environment, thus they are not strongly correlated. As a consequence observed stellar streams cannot directly be converted to the distribution of the DM debris. Furthermore, it shows the importance of cosmological initial conditions as well as the realistic structure of satellite galaxies in determining the final distribution of the satellite streams around the MW.

\subsection{Discussion}

The fact that the debris (particularly, the stellar debris) has some degree of flatness shows that with a cosmologically motivated initial setup (like, in our case, from Aq simulations data) it is possible to obtain a flat spatial distribution of debris. On the other end, our debris does not show any distribution on a unique plane, as can be seen from Figure \ref{fig:c_mollweide_all}, where within different radii the debris tensor seems to occupy different regions in the angles-space in the same simulation and for each individual component (DM and stars). This contrasts with what is stated by \cite{Pawlowski12}, that goes in favor of the formation of a plane of debris. Furthermore, since our results predict no systematic orientation of the stellar and DM debris, this means that other techniques may be needed to trace the distribution of DM streams, other than addressing the distribution of the stellar streams alone. 

Regarding the mass loss of the satellites in our simulations, S19 stated that tidal stripping affects the luminosity but not the metallicity of the stellar populations of satellites, and since the luminosity-metallicity relation is satisfied for the observed UFDs, then the stripping of stars acting on satellites must be moderate. This conclusion is in line with what we got for the stripping fraction of stars, which is not high for stellar satellite matter.
S19 found the result in agreement with previous estimates such as from \cite{Kirby13}.

When considering the fraction of surviving satellites, regardless of whether the satellites are DM-only or hybrid, this is not the feature that determines their survival. In fact, no systematic differences were found between the MJ16 simulational sample and our simulational sample.
The inner properties of the satellites, as previously stated, differ in the shape of their density profiles, shallower for DM in hybrid satellites than in DM-only satellites. However, the difference in slope of these profiles is limited, and the energy distribution of DM particles is similar for DM-only and hybrid satellites.
Instead, it seems that other drivers, such as the satellite initial conditions of our simulations, determine the survival probability of the satellites, therefore this is related to their orbital parameters and intrinsic structure.

The fact that the tidal stripping exerted by the MW is not strongly efficient on its satellite galaxies seems to be compatible with what \cite{Penarrubia08} found in their simulations. In particular, they found that even after $99\%$ of the stars get stripped from a satellite, the King profile \citep{King66} of the remaining stellar component is kept unchanged. Furthermore, they found that the stripping of stars is efficient only when the satellites approach their orbital periapsis, i.e., when they get closer to the MW centre. 
Previous work from \cite{BullockJohnston05} that has investigated the effect of MW-satellites interactions in the past epochs has underlined the importance of these mergers to form the stellar halo. However, they adopted a combination of N-body methods and semianalytical models to address their study. The MW, for instance, was added only analytically as a time-evolving potential. Here we have taken advantage of full N-body simulations, and the live evolution of the gravitational potential of the MW, with effects such as dynamical friction \citep{Chandrasekhar43} being naturally incorporated, however without a secular long-term growth. 
Furthermore, our analysis is different from the one of \cite{BullockJohnston05} in the sense that it focuses on the current distribution of the residual debris coming from recently accreted satellite structures, rather than focusing on the build-up of the MW halo from past accretion processes.

\begin{acknowledgements}

Funded by the Deutsche Forschungsgemeinschaft (DFG, German Research 
Foundation) -- Project-ID 138713538 -- SFB 881 (``The Milky Way 
System'', subproject A02). The authors acknowledge support by the state of Baden-W\"urttemberg through bwHPC. 
Part of this research was carried out on the High Performance Computing resources at New York University Abu Dhabi.
We thank J. Frings for providing data for the selection of dwarf galaxies employed for the simulations. 
M.M. thanks V. Springel and his collaborators for support using the code \texttt{Gadget-4}. 
M.M. thanks also M. Arca Sedda, A. Borch, A. Pasquali, B. Avramov, B. Bidaran, A. Savino, T. M. Jackson and M. Donnari for further discussion. 

\end{acknowledgements}


\bibliographystyle{aa} 

\bibliography{main}


\end{document}